**Giant mobility of surface-trapped ionic charges following liquid tribocharging**

Zouhir Benrahla[1], Tristan Saide[1], Louis Burnaz[1], Emilie Verneuil[1],
Simon Gravelle[2], Jean Comtet[1*]

*[1] Soft Matter Sciences and Engineering, CNRS, ESPCI Paris, PSL University, Sorbonne Université, 75005 Paris, France*
*[2] Univ. Grenoble Alpes, CNRS, LIPhy, 38000 Grenoble, France*

* jean.comtet@espci.fr

**The sliding motion of aqueous droplets on hydrohobic surfaces leads to charge separation at the trailing edge, with implications from triple-line friction to hydrovoltaic energy generation. Charges deposited on the solid surface have been attributed to ions or electrons ripped off from the liquid drop. However, the dynamics and exact physicochemical nature of these surface-trapped charges remains poorly explored. Here, we take advantage of a scanning-based electrostatic mapping technique, to directly quantify the spatiotemporal dynamics of surface deposited charges in the wake of droplets sliding on hydrophobic surfaces. We confirm the ionic nature of these interfacially trapped charges, and evidence that they undergo very fast bidimensional diffusive transport, gliding with low friction at the solid/gas interface. We interpret our observations in the framework of molecular dynamics simulation of hydrated ions adsorbed on solid surfaces, revealing a peculiar transport mechanism limited by purely interfacial friction of the ionic solvation shell with the solid surface. By uncovering the unexpected dynamics of these ionic puddles - a distinct state of interfacial ionic matter - our findings have general implications for molecular-scale ionic transport, electrified matter and wetting dynamics at interfaces.**

Tribocharging - the process by which surfaces charge following frictional contact - is a long standing issue in physics and material science[1], with broad implications ranging from the mundane experience of "static electricity"[2] to protoplanetary formation from spatial dust aggregation[3]. Amongst the diverse electrification processes which usually involve contacts between solid-state insulating or conducting materials, *liquid tribocharging* stands aside, with electrification occurring through the sliding motion of a liquid water droplet on a solid hydrophobic surface. In this peculiar situation, surface-deposited charges are thought to originate from bulk ionic carriers present in the liquid drop[4,5], while the gliding motion at the deformable trailing contact line couples to the charge deposition processes, leading to new dissipation pathways[6–8]. Liquid tribocharging has thus raised considerable interests in the recent years[9,10], being commonly experienced in a variety of situations of applied interests, such as pipetting[11,12], self-charging of spray[13], droplet jumping[14] and directional droplet transport[15], as well as in the development of nanogenerators with promising performances[16,17].

At a mechanistic level, charging is thought to stem from ions adsorbed specifically on the solid/liquid interface inside the droplet, which are deposited on the solid surface due to their sluggish dynamics during droplet dewetting[4,5]. Interestingly, electrons have also been put forward as possible charge-carriers[18–20]. However, due to the challenges associated with accessing interface-specific information at appropriate spatial and temporal resolution, the nature, physicochemical state and dynamics of the charges deposited on the surface of the solid has remained largely unexplored. In particular, most studies approaching these questions focused either on the droplet charge or spatially-averaged surface charge (except for an early seminal work[21]), therefore providing a partial and incomplete picture of the deposited charges dynamics.

Here, we harvest an in-situ scanning based potential mapping approach, to reveal the spatial and temporal evolution of surface-deposited charges following the sliding of droplets on hydrophobic surfaces. We first evidence that surface charging at low droplet numbers occurs in a heterogeneous fashion with patches of alternating signs observed down the droplet pathway. At large droplet numbers, we observe a monotonously decreasing potential, whose sign and magnitude can be reversibly tuned by the bulk droplet pH. The deposited charge follows a similar trend with pH as the surface charge at the interface between non-ionizable hydrophobic surfaces and water, suggesting the involvement of physisorbing surface-active ionic species (such as water self-ion and dissolved $CO_2$) in the surface charging mechanism. Remarkably, the charges initially deposited along the wake of the droplet tend to spread apart on the solid surface over time, consistent with a bidimensional diffusion process associated with ultra-high 2D gliding mobility exceeding that of standard salt in bulk water. Transport is furthermore found to be independent of the surface charge density, indicating that charge dynamics is dominated by surface interactions rather than inter-charge repulsions. We interpret these experimental observations through MD simulations of hydrated ions adsorbed on model hydrophobic surfaces, evidencing that interfacial mobility in this peculiar situation becomes uncorrelated to bulk ionic diffusion, but leads to the emergence of an interface-specific dissipative process set by the interaction of the ionic hydration layer with the surface of the solid. Our findings have broad implications and open up many exciting avenues, from the role of water and ionic adsorbates in solid-solid electrification, to the fundamentals raised by the dynamics of these *ionic puddles* – a new state of interfacial ionic matter.

## Results and discussion

***Surface potential mapping reveals heterogeneous charge deposition by sliding droplets.*** To probe the spatial and temporal dynamics of surface-deposited charges following droplet sliding, we take advantage of a scanning-based electrostatic potential mapping technique, pictured in Fig. 1 and detailed in SI.S2. Briefly, the aqueous droplets impact and slide on a tilted hydrophobic surface, consisting of a glass plate grafted with a self-assembled monolayer of long $C_{18}$ chains (Fig. 1C-D, and SI.S1). By sliding on such a hydrophobic surface, droplets acquire a net positive charge[4–6] which is readily detected through the measurement of their discharge current on a downstream electrode (Fig. S2), leading presumably to the deposition of charges of opposite signs on the solid surface. The deposited surface charges of density $\sigma_S$ [C.m$^{-2}$] lead to the build-up of a surface potential $V_S$ [V], with $\sigma_S = (\varepsilon_r \varepsilon_0 V_S)/h$, where $\epsilon_r \approx 4.8$ [-] the relative dielectric constant of the glass, $\epsilon_0$ [F.m$^{-1}$] the vacuum permittivity and $h$ [m] the glass plate thickness. This surface potential can be spatially mapped by a non-contact field-compensating electrostatic voltmeter with a metallic electrode (diameter 790 µm) oscillating above the surface. In this geometry, the probe oscillation induces an AC capacitive current $i_{AC}$ proportional to the probe/surface potential difference $V_{\mathrm{probe}} - V_S$, which is nullified to access the local surface potential $V_S$ and associated surface charge density $\sigma_S$. By scanning the electrostatic voltmeter above the surface, surface potential maps are readily obtained.

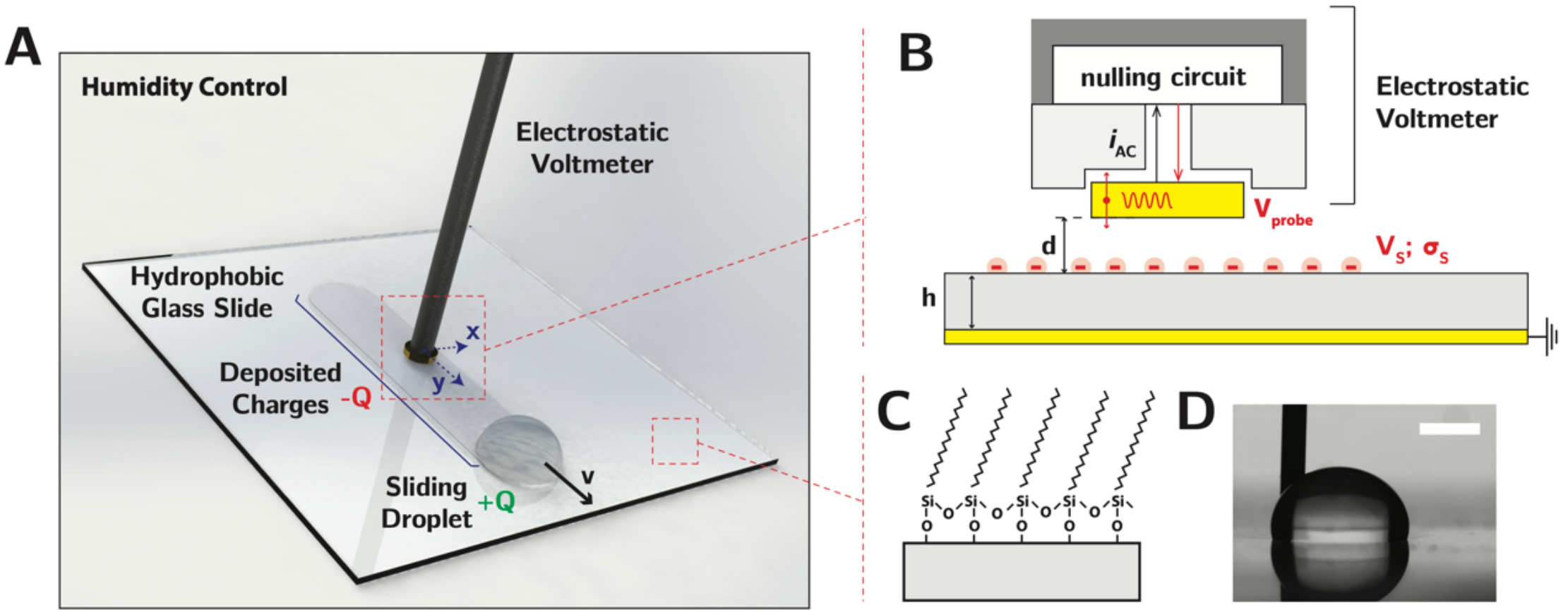

**Figure 1. Large-scale surface potential mapping. (A)** Schematic of the set-up used for the spatio-temporal characterization of surface charging by sliding droplets. A droplet of liquid impacts and slides with velocity $v \approx 0.5$ m.s$^{-1}$ over a tilted hydrophobic glass surface. The local density of surface deposited charges $\sigma_S$ [C.m$^{-2}$] induces a surface potential $V_S$ [V] which is spatially mapped by a scanning electrostatic voltmeter. **(B)** Schematic principle for surface potential mapping using the electrostatic voltmeter. Surface deposited charges with density $\sigma_S$ are represented in red and induce a surface potential $V_S$. A metallic probe (potential $V_{\mathrm{probe}}$) oscillates at a distance $d = 700$ µm from the solid surface, inducing an AC current, proportional to the voltage difference $V_S - V_{\mathrm{probe}}$. A nulling circuit adjusts the probe potential so that $V_S = V_{\mathrm{probe}}$. The hydrophobic glass plate holding the deposited charges has a thickness $h = 2$ mm and is grounded on the backside (yellow electrode). **(C)** Schematic of the surface chemistry, with $C_{18}$ chains grafted to the glass surface, rendering the surface hydrophobic (see SI.S1 for details on substrate preparation). **(D)** Image of a water droplet sitting on the surface. Scale bar is 2 mm.

We first investigate in Fig. 2A how the local surface charge state varies upon increasing the number of sliding droplets. We systematically start our experiments by annealing surfaces at 150°C, allowing us to obtain an initially homogeneous surface state free of residual surface charges, as shown in Fig. 2A, *i,* where the RMS voltage is lower than 2 V. Following sliding of the first droplet (Fig. 2A, *ii)*, we clearly evidence the deposition of charges in the wake of the droplet trajectory (vertical dashed line). Unexpectedly, while previous ensemble measurements highlighted a systematically positive charging of the droplet, interpreted as the deposition of purely negative charges onto the surface[4,6,22], our local measurements evidence that charges deposition occurs through patches of alternating signs. Starting with the deposition of negative charges at the impact point (top of the image, red region), the deposited charge shows clear oscillations along the vertical y-axis reaching successively positive (green) and negative (red) values. As schematically represented in Fig. 2B, such alternating charge patches might stem from the evolution of the droplet charge during sliding: following impact and initial sliding, the drop becomes positively charged and its potential rises, which ultimately shifts the sign of the deposited charges to positive values, etc. Upon increasing the number of sliding droplets (Fig. 2A, *ii* to *iv*) positively charged domains in green are progressively smeared out. At the same time, the surface becomes increasingly negatively charged until reaching uniformly negative values (Fig. 2A, *v* and *vi*), a clear signature of the preferred transfer of negative charges from the solid/liquid to the solid/gas interface.

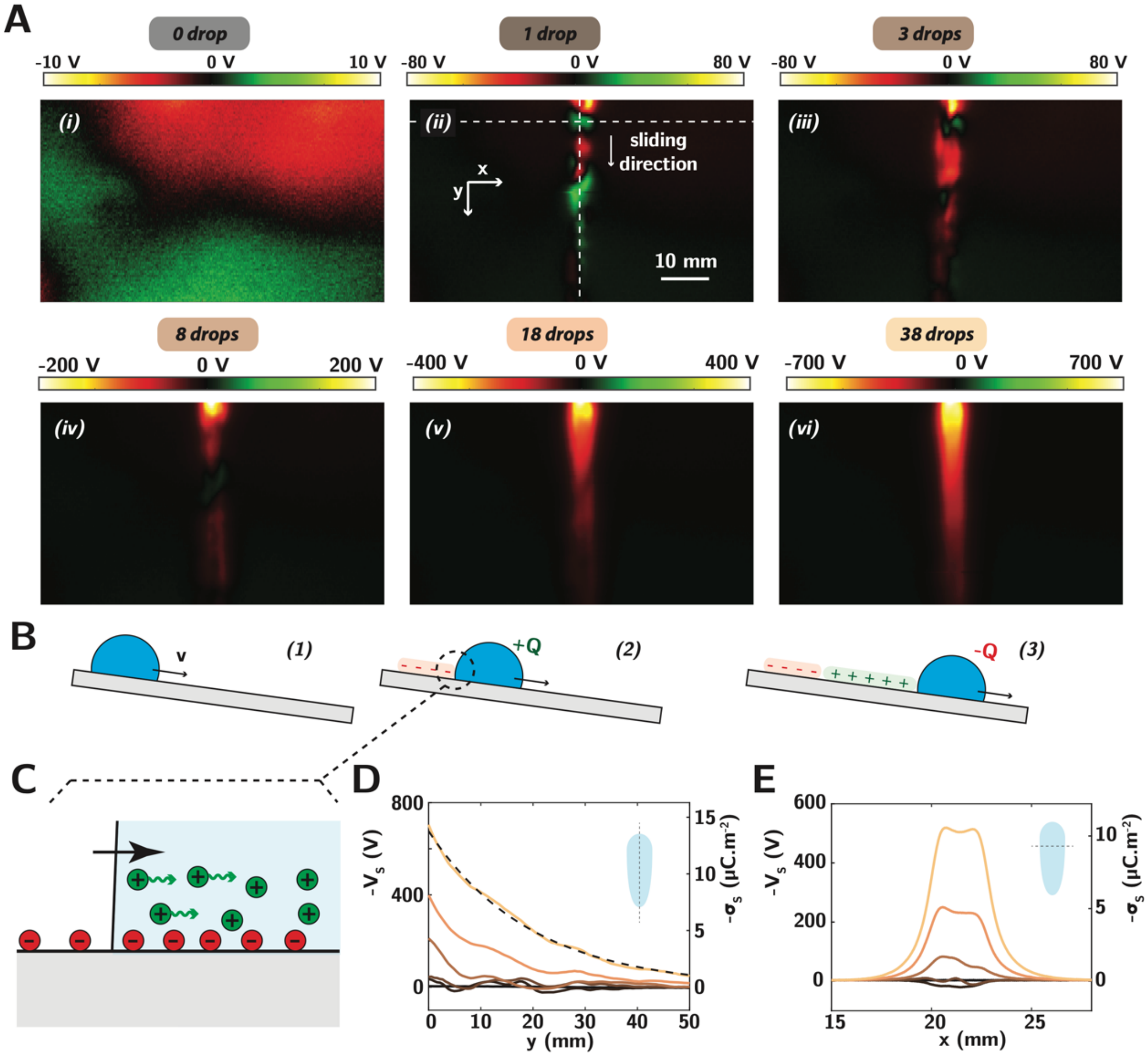

**Figure 2. Heterogeneous spatial surface charging following liquid triboelectrification. (A)** Evolution of the surface potential for increasing droplet number. The point of impact of the drop is situated at the top of the image. The interval time between each scan is of the order of 20 minutes and the droplets are launched sequentially on the same path at a rate of $\approx 1$ droplet per second. Droplets are composed of deionized water of pH ~5.5. **(B)** Schematic of the observed charge oscillation during droplet sliding. The droplet is initially neutral and gets positively charged ($+Q$, green) while depositing negative charges (red) to the surface. Upon accumulation of excess positive charges in the drop, positive charges (green) are transferred to the surface, and the drop charge relaxes to negative values, leading to the observed charge

oscillations. **(C)** Zoom-in at the triple-line on the rear of the droplet. Inside of the droplet, negative charges (red) adsorbed to the interface are screened by counter-ions in the liquid (green). Upon dewetting of the triple line, negative charges adsorbed to the solid/liquid interface are transferred to the solid/gas interface. **(D, E)** Spatial evolution of the surface potential along (D) and perpendicular (E) to the drop sliding path, on a surface charged from 0 to 38 drops (under the same conditions as in (A)). The dashed line in D is an exponential fit. Insets are top views of the droplet patterns.

A coarse mechanism for the charging process is schematically represented in Fig. 2C. Inside the droplet, the solid/liquid interface bears a negative charge originating from specific ionic adsorption (red negative charges), which are screened by counterions in the liquid (green positive charges) over nanometric distances[23]. Upon displacement of the triple-line, counter-ions are convected by the liquid, while some surface-adsorbed ions are transferred from the solid/liquid interface to the solid/gas interface, due to their sluggish dynamics induced by their interactions with the surface. It thus appears interesting to compare the charge densities deposited on the solid surface with those generated through ionic adsorption in the solid. The surface potential at the impact point is found to be of the order of 30 V, corresponding to surface charge density of $\approx 0.5\ \mu\mathrm{C.\,m^{-2}}$, several orders of magnitude smaller than the surface charge density of 0.1-1 $\mathrm{mC.m^{-2}}$ expected to develop at similar hydrophobic surface in deionized water[24], demonstrating the very low probability associated with this charge ripping process. Conversely, at large drop numbers, the absolute surface potential can reach up to 1800 V at the impact point (the saturation limit of our instrument) associated with absolute charge densities up to 35 $\mu\mathrm{C.\,m^{-2}}$. In this limit, the potential decays approximately exponentially along the drop path, with a characteristic distance of $\approx 20$ mm (Fig. 2D, black dashed line), a feature reminiscent of previous observations of the charge of the droplets that was found to saturate with the distance they were allowed to slide[4,6]. In the direction transverse to the sliding path (Fig. 2E), the surface potential is approximately constant along the wake of the drop, demonstrating the occurrence of homogeneous charging, although some slight over-charging at the edges of the wake can be observed far from the impact point (Fig. 2F).

***Surface trapped charges glide with ultra-low friction on the solid surface.*** Observing this peculiar trapping process raises the question of the dynamics of the surface deposited charges. We thus track the long-term spatial and temporal evolution of the surface potential profile following charging of the surface by a large number of droplets (SI Fig. S1). As shown in Fig. 3A, the charges initially deposited and confined along the drop wake are found to spread out on the surface over time, suggesting their large mobility through diffusion over several hours. We highlight that such spatio-temporal evolution is not affected by the scanning rate of the electrostatic probe, indicating that this phenomenon is an intrinsic surface relaxation process. To further quantify this transport process, we report in Fig. 3B, the temporal evolution of the surface potential profile perpendicular to the drop wake close to the impact point, evidencing again the spreading of the deposited charge. Due to the geometry of the deposited charge patch, surface diffusion can be well-described by a 1D transport process (SI. S3) and quantified by evaluating the temporal evolution of the Full Width at Half Maximum $\Delta$ of each profile. As reported in Fig. 3C, following an initial sublinear evolution associated with the finite width of the deposited charge patch, $\Delta^2$ shows a clear linear scaling with time, consistent with a purely diffusive process characterized here by a surface diffusion coefficient $D_{\text{surface}} \approx 2.6 \cdot 10^{-9}$ m$^2$.s$^{-1}$. This linear scaling is obtained while the voltage at the center of the trail decreases over one order of magnitude, evidencing that the charge transport mechanism is independent on the charge density. This independence indicates the negligible contribution of the repulsive electrostatic interactions between the deposited charges to the transport process. For surface charge densities $\sigma \approx 35$ μC. m$^{-2}$, the typical distance between charges reaches $l \sim (\sigma/e)^{-1/2} \approx 70$ nm, which is indeed slightly larger than the Bjerrum length $l_B \sim e^2/(4\pi\epsilon_0 k_B T) \approx 57$ nm evaluated in air (with $\epsilon_r \approx 1$), which defines the typical distance at which thermal motion overcomes electrostatic interactions.

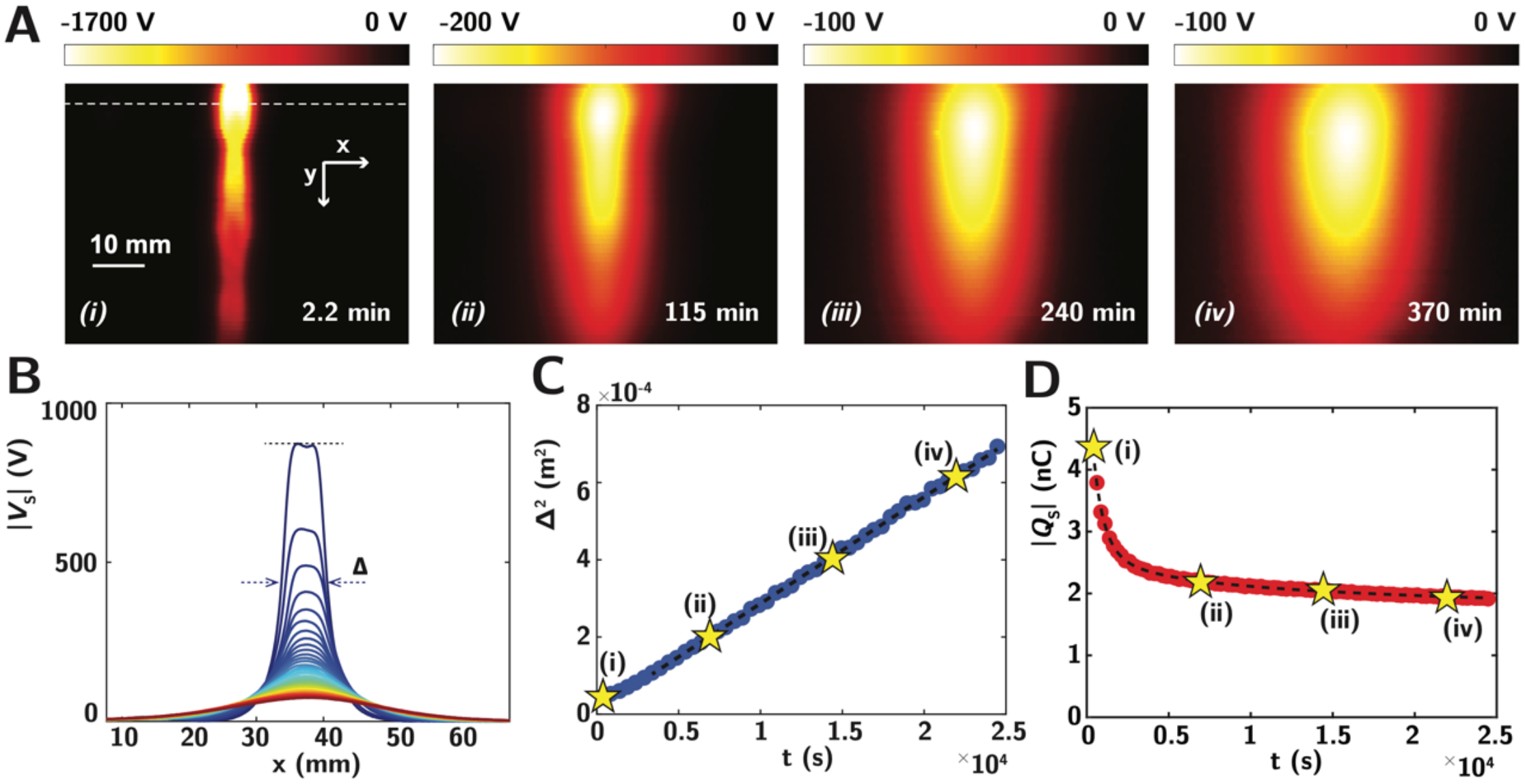

**Figure 3. Surface charge diffusion. (A)** Temporal evolution of a patch of surface charges, showing its progressive spreading on the surface of the solid over increasing time. The initial patch is formed by sliding 35 droplets (deionized water, pH ~5.5) at a rate of ≈ 1 droplet/second. **(B)** Temporal evolution of the surface potential transverse to the sliding droplet path, characterizing the Full Width at Half Maximum $\Delta$ for the initial profile in dark blue. The time-interval between each curve is 502 s and the elapsed time between the initial and final profiles is $2.5 \cdot 10^4$ s ≈ 7 hours. **(C)** Evolution of $\Delta^2$ as a function of time. **(D)** Relaxation of the total surface deposited charge $Q_S$ over time. The dashed line is an empirical fit combining exponential and power-law relaxation (see Fig. S5 for details). In (C) and (D), the stars indicate the timestamps corresponding to the four images in (A).

Such diffusion process was evidenced on distinct surfaces, showing a systematically linear variations of $\Delta^2$ with time. The associated diffusion coefficient exhibited slight variations from sample to sample - consistent with variations in the microscopic surface state - yielding diffusion coefficients ranging from $1 \cdot 10^{-9}$ to $7 \cdot 10^{-9}$ m$^2$.s$^{-1}$ (Fig. S7). Surprisingly, such values are found to be *of the same order or* even *larger* than the typical diffusion coefficient of simple ions in bulk water, taking as a reference a diffusion coefficient $D_{\mathrm{Na+}} \approx 1.4 \cdot 10^{-9}$ m$^2$.s$^{-1}$ for sodium ions at 25°C[(25)]. The remarkably high mobilities we measured are unexpected, as wall-induced friction is typically expected to slow down the dynamics compared to liquid friction in bulk. A mechanism consistent with the high mobility of the ions is proposed below.

Another remarkable aspect of this charge transport process lies in the exceptionally large stability of the deposited surface charge, as demonstrated in Fig. 3D, where we present the temporal evolution of the total charge deposited on the surface $Q_{\mathrm{S}}$, obtained by integrating the total charge density of the imaging area. We observe in this figure a short-term exponential-like relaxation with a characteristic time of 15 to 20 minutes, after which the charge transitions to a much slower relaxation process (see Fig. S5 for details). We propose that this transient relaxation of the surface potential is due to the equilibration of the physicochemical state of the deposited charge, possibly caused by the progressive build-up of a physisorbed water shell around the ions, which may screen the induced field and surface potential, or by the evaporation of a fraction of surface trapped ions (anticipating our simulation results below). Other irreversible charge relaxation pathways out of the surface could occur such as the attraction of oppositely charged species from the atmosphere or the leakage of surface-deposited charges through the sample to the grounded electrode. The long remanence and very slow relaxation evidenced on larger time-scales after $0.5 \cdot 10^4$ s demonstrates the reduced pathways for such first-order charge annihilation processes and characterizes again the purely interfacial bidimensional nature of the transport evidenced here.

***Water pH can tune surface charging.*** These peculiar observations raise the question of the physicochemical state of these surface deposited charges and their relation with ionic adsorption at the solid/liquid interface close to the dewetting contact line. To probe this question in more detail, we investigate the effect of water ionic content on charge deposition. We probed in particular the effect of pH, which is known to have a major influence on the surface charge at solid/water interfaces. Note that we used here solutions with a background level of 10 mM salt, in order to keep the Debye screening length constant over the entire range of studied pH.

As evidenced in Fig. 4A, we observe a dramatic effect of pH on surface charging, with a switch in the sign of the deposited surface charge when decreasing the pH to acidic values. This trend is confirmed in Fig. 4B, where we report the evolution of the total surface charge $Q_S$ [C] as a function of the droplet pH, showing a reversal from positive to negative charging around pH 3.5 (comparing green and red regions). Remarkably, the trend observed here of a monotonic decrease of the value of the surface charge from pH 2 to pH 7, followed by a plateau up to pH 9 is consistent with surface (or zeta) potential evolution typically evidenced in water on various hydrophobic surface[26], pointing to the occurrence of a similar charging mechanism. Surprisingly, while this general trend has been reported in a robust manner on a variety of hydrophobic surfaces, the exact underlying mechanism remains unclear[27]. Original interpretations pointed to the role of the specific interfacial adsorption of the water self-ions $\mathrm{H_3O^+}$ and $\mathrm{OH^-}$ on surface charging at respectively low and high pH values[28–30]. Concurrent explanation involves the specific adsorption of hydrogenocarbonate ions $\mathrm{HCO_3^-}$, originating from $CO_2$ dissolution at neutral to basic pH[31], as well as surface-specific impurities[32,33] or reorganization in the interfacial hydrogen-bonding network of water[34–36]. In our case, we observe a decrease of the magnitude of the deposited charge when reaching large pH $\geq 10$ (Fig. 4B), a range which has been explored much less systematically for these interfaces. As reported in previous studies[31], such decrease in the absolute value of the charge at high pH could be a signature of the important role played by ions originating from dissolved gaseous $CO_2$. Indeed, this decrease coincides approximately with the pKa of 10.3 for the equilibrium between $HCO_3^-$ and $CO_3^{2-}$ [31], shown by the vertical dashed line in Fig. 4B. Given that hydrophobic interfaces are typically attractive to $HCO_3^-$ and repulsive to $CO_3^{2-}$ [31,32], the observed trend would then be a signature of the adsorption of $HCO_3^-$ to the interface over this intermediate pH range. This hypothesis would however still require ionic adsorption energies at these solid/hydrophobic interfaces to reach larger values than that measured at solid/air interfaces by surface tension measurements[37]. All-in-all, while the debate for the exact physicochemical mechanism for hydrophobic surface charging remains a lively one, the most probable candidates here are thus expected to be surface-active physisorbing ions present in solution and whose relative concentrations would be largely affected by the pH, namely $H_3O^+$, $OH^-$ and potentially $HCO_3^-$ ions.

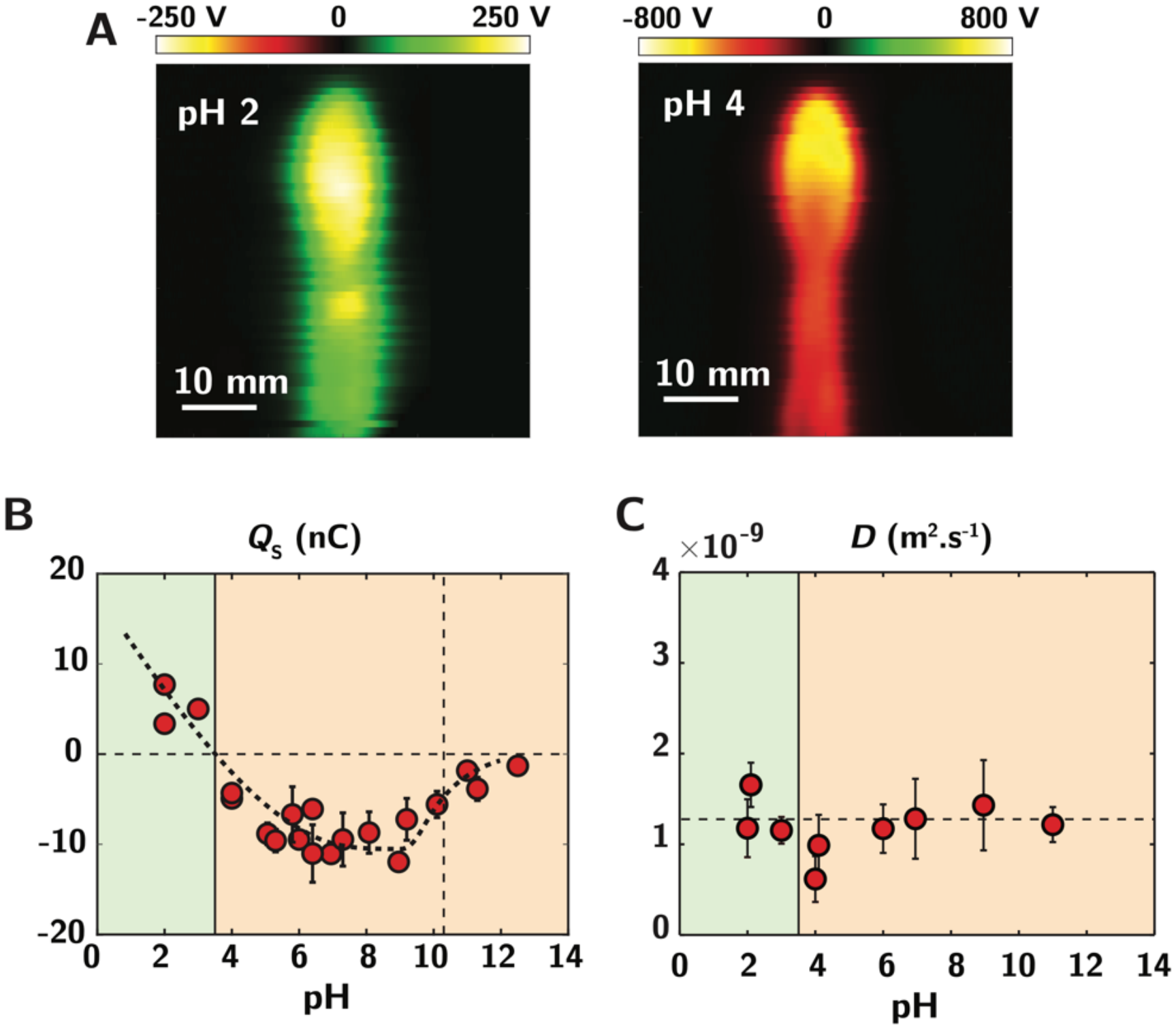

**Figure 4. Physicochemical control of charge deposition and surface diffusion. (A)** Surface potential maps obtained through successive droplet sliding at pH 2 and pH 4, showing reversal of the deposited charges. pH is adjusted in a solution with 10 mM KCl background salt (see SI.S4). **(B)** Evolution of the surface charge $Q_S$ with droplet pH. The dashed line is a guide to the eyes. The green and red domains correspond respectively to positive and negative charges, with the vertical solid line at pH ≈ 3.5 indicating the sign reversal of the deposited charge. The vertical dashed line indicates the pKa ≈ 10.3 of the $HCO_3^-/CO_3^{2-}$ couple. **(C)** Surface charge diffusion coefficient $D$ as a function of pH is constant for a given surface within experimental error.

We might expect the presence of these distinct species to affect interfacial mobility. In bulk water at 25°C, the mobilities of these various ions can indeed vary by a factor of 10 from $9.4 \cdot 10^{-9}$ m$^2$.s$^{-1}$ and $5.2 \cdot 10^{-9}$ m$^2$.s$^{-1}$ for $\mathrm{H_3O^+}$ and $\mathrm{OH^-}$ [38] down to $1.2 \cdot 10^{-9}$ m$^2$.s$^{-1}$ for $\mathrm{HCO_3^-}$ [39]. These differences are attributable to the peculiar mode of diffusion for $\mathrm{H_3O^+}$ and $\mathrm{OH^-}$, which does not involve Fickian-like drag of the ionic center of mass, but rather involves Grotthus hopping and tunneling of the excess proton from water molecules to water molecules leading to ultrafast transport. Remarkably, such large modulation of the ionic mobility *at the interface* is not recovered in our experiments, with a diffusion coefficient measured to be constant within experimental error over the whole range of pH values between 2 and 12 (Fig. 4C). These two observations of *(i)* a diffusion process independent of the exact ionic nature and charge density and *(ii)* mobility values which can exceed standard bulk ionic diffusion coefficient for simple ions suggest the occurrence of a peculiar mode for 2D ionic transport at solid/gas interfaces, which we will address in the following discussion.

***Ionic transport in the limit of interfacial hydrated friction.*** A question arises with respect to the physicochemical state of these surface adsorbed ions. Although trapped at the solid/gas interface, we might expect these ions to still be surrounded by a few water molecules, due to the high enthalpic cost of 50 $k_BT$ needed to fully strip away their water hydration shell[4,40]. With this picture in mind, we turn to molecular dynamics simulation in Fig. 5 to get more insights into the possible molecular transport mechanisms for these mobile surface-adsorbed ions. As shown in Fig. 5A, we consider the dynamics of a single ion adsorbed on a hydrophobic self-assembled monolayer surface (here a bicarbonate $\mathrm{HCO_3^-}$ ion, with $\mathrm{OH^-}$ and $\mathrm{H_3O^+}$ ions also considered in our simulations). We consider these ions under distinct hydration levels, characterized by the number $N_\mathrm{w}$ of water molecules adsorbed around them, with here respectively $N_\mathrm{w} = 5$ and 161 water molecules, a range justified by Grand Canonical simulations at various relative humidities. Note that in these simulations, the ionic nature was found to affect only weakly the associated hydration level (Fig. S9).

As shown in SI Movie S1-2, we let these hydrated ions evolve spontaneously at the interface, evidencing a random diffusive-like surface motion (see individual trajectories in Fig. 5A-B). At a qualitative level, we already evidence significant difference in charge mobility for distinct hydration level, with the largest nanodroplets displaying slower dynamics (comparing trajectories shown in the center plot in Figs. 5A and B). Interestingly, while the motion for the large ionic nanodroplet is purely confined in 2D at the interface (B, left panel), for the weakly hydrated ion, we evidenced the occurrence of transient detachments from the surface (A, left panel showing a 3D trajectory), an effect we will return to later.

To characterize these differences of dynamics in more quantitative terms, we compute in Fig. 5C the temporal evolution of the Mean-Squared Displacement, restricting our analysis to the diffusive dynamics in the adsorbed state, from which the associated 2D diffusion coefficient $D_\mathrm{ion}$ [m$^2$.s$^{-1}$] is obtained as $\langle (x(t) - x_{t=0})^2 \rangle = 4D_\mathrm{ion}t$. The associated to diffusion coefficients span over one order of magnitude, from $D_\mathrm{ion} \approx 2.5 \cdot 10^{-8}$ m$^2$.s$^{-1}$ to $D_\mathrm{ion} \approx 3 \cdot 10^{-9}$ m$^2$.s$^{-1}$, when the number of water molecules increases from 5 to 161 (Figs. 5A-B). A first comment arises regarding the order of magnitude of these associated diffusion coefficients, which are found to be up to one order of magnitude larger than in bulk water, highlighting the very small friction experienced by these surface physisorbed hydrated ions.

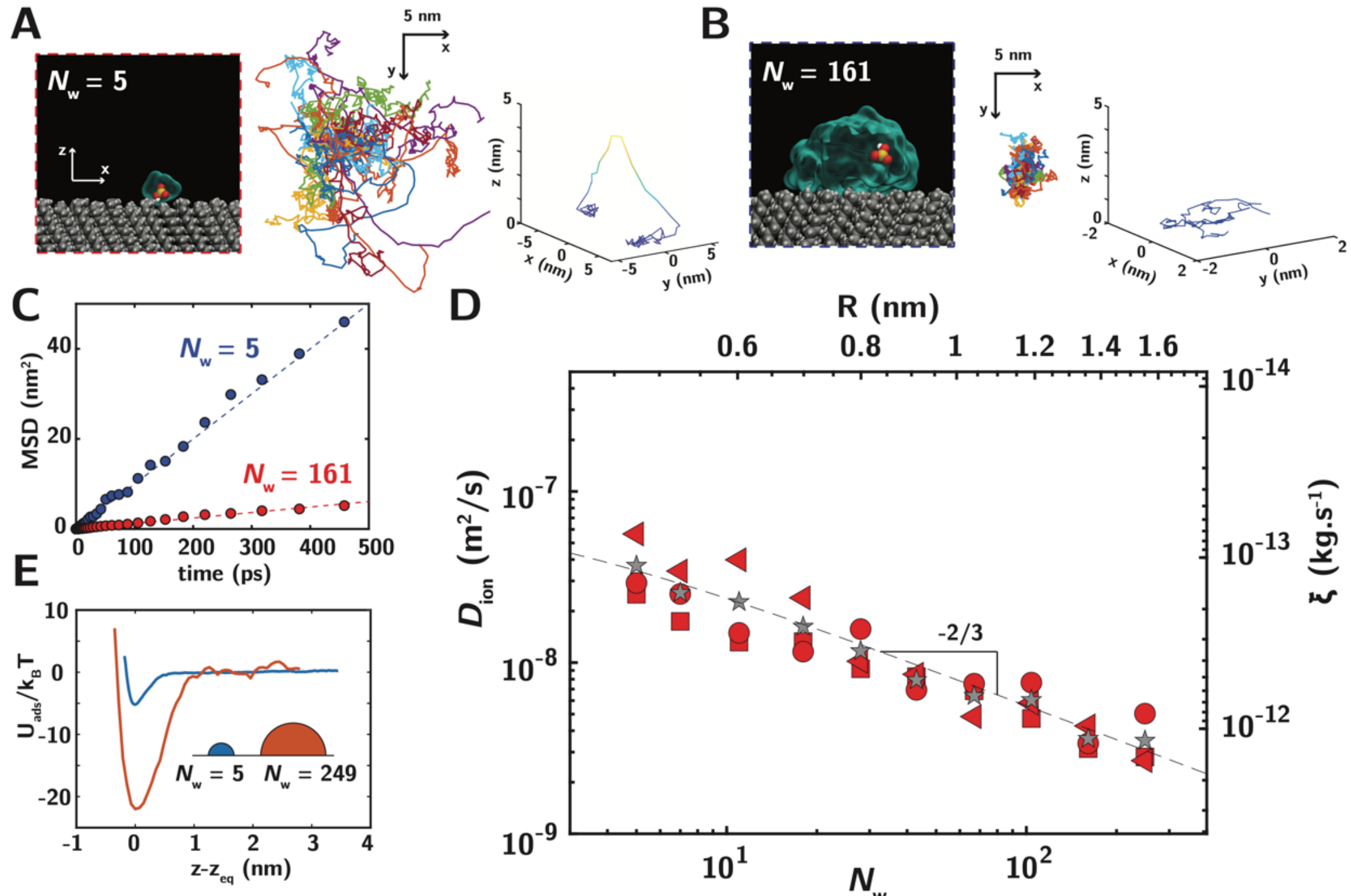

**Figure 5. Molecular insight into interfacial ionic diffusion. (A-B)** Snapshot of MD simulations for a surface adsorbed ion surrounded by hydration shell with $N_{\mathrm{w}} = 5$ and 161 water molecules (left panel), along with typical 0.5 ns trajectories, projected on the (xy) plane (center panel) and examples of single trajectories in (xyz), with the vertical z position color-coded according to height from blue to yellow (right panel), showing the occurrence of transient detachment out of the surface for $N_{\mathrm{w}} = 5$ in (A). **(C)** Effective Mean-Squared Displacement (MSD) in the adsorbed state as a function of time for the two equilibrium states depicted in (A) (averaged over 30 independent simulations). **(D)** Evolution of diffusion coefficient in the absorbed state with the number of physisorbed water molecules. The dashed line represents a power-law scaling of -2/3, and the different symbols correspond respectively to $OH^-$ (triangles), $H_3O^+$ (circles) and $HCO_3^-$ (squares). Grey stars indicate the averaged diffusion coefficient for these three species. **(E)** Interfacial physisorption energy $U_{\mathrm{ads}}/k_B T$ for a bicarbonate ion surrounded by 5 and 249 water molecules. $z_{\mathrm{eq}}$ characterizes the equilibrium position at which $U_{\mathrm{ads}}$ is minimal.

To analyze these effects in more detail and evidence the limiting transport steps in relation to ionic hydration, we plot in Fig. 5D the evolution of the diffusion coefficient as a function of the number of water molecules $N_{\mathrm{w}}$. The diffusion data is shown for the three types of investigated ions, with $OH^-$, $H_3O^+$ and $HCO_3^-$ corresponding respectively to the triangle, circle and square symbols in red. Since the diffusion coefficients appear to be weakly dependent on the ionic nature but are mostly determined by the hydration number $N_{\mathrm{w}}$, we further characterize the average diffusion coefficient across these three species for the various $N_{\mathrm{w}}$ using grey stars. Interestingly, the surface diffusion coefficient $D_{ion}$ follows a clear power-law decrease with $N_{\mathrm{w}}$, associated with the approximate scaling $D_{\mathrm{ion}} \sim N_{\mathrm{w}}^{-2/3}$, shown by the dotted line in Fig. 5D. The approximate surface and volume of the drop scales respectively as $N_w^{2/3}$ and $N_w$. Hence, the scaling $D_{\mathrm{ion}} \sim N_{\mathrm{w}}^{-2/3}$ suggests an interface limited transport process, reminiscent of previous experimental[41] and computational studies on various adsorbate systems[42,43].

Building upon this idea of interfacially-limited transport, we write the diffusion coefficient as $D_{\mathrm{ion}} = kT/\xi_{\mathrm{ion}}$ with $\xi_{\mathrm{ion}}$ [kg.s$^{-1}$] the friction coefficient of the hydrated ion, evolving within the range $10^{-13}$ to $10^{-12}$ kg.s$^{-1}$ (Fig. 5D, right y-axis). Following up on the observed scaling of $D_{\mathrm{ion}}$ as $N_{\mathrm{w}}^{2/3}$ in Fig. 5D, we express the ionic friction coefficient as $\xi_{\mathrm{ion}} = A \cdot \lambda$, the product of the nanodroplet surface area $A$ [m] by an interfacial friction coefficient $\lambda \approx 1.8 \cdot 10^{5}$ Pa.s.m$^{-1}$, where A is geometrically determined from $N_w$ (SI. S5). This interfacial friction coefficient is expected to depend on the specific interactions between the last molecular layer of liquid and the solid surface. For a situation where water would flow alongside the same hydrophobic surface, this interfacial friction $\lambda$ would lead to a slip length denoted $b$ and expressed as $b = \eta/\lambda$ with $\eta$ the bulk viscosity. This situation was independently simulated through a Couette flow, as detailed in SI.S5, and we obtain a simulated value $\lambda^{c} = \eta/b_{\mathrm{c}} \approx 6 \cdot 10^{5}$ Pa.s.m$^{-1}$ with $b_c = 1.6$ nm the slip length measured at the OTS/water interface. This value of the interfacial hydrodynamic friction coefficient is of the same order that obtained for the nanodroplets. This agreement validates that the limiting process for the diffusion of the hydrated ions on the surface arises primarily from the interfacial friction of the ionic solvation shell with the surface. Note that this scaling is expected to break down in the limit of ultra-low slip length for which the droplets might adopt a rolling-like motion, opening additional volumic dissipation channels, related to bulk nanodroplet deformation[44].

Returning to the trajectories shown in Figs. 5AB, our simulations evidence a second unexpected phenomenon at play for very small nanodroplets, whereby the solvated ions can leave the surface, with an effective detachment rate which decrease for increasing droplet size (Fig. S8). To characterize the possibility for such thermally activated surface desorption, we quantify in Fig. 5D the effective adsorption energy of ~~the~~ solvated ions on the solid surface for solvation shells consisting of 5 and 249 water molecules, evidencing a typical adsorption energy $U_{\mathrm{ads}}$ of 4 and 22 $k_{\mathrm{B}}T$ respectively. Assuming a molecular attempt frequency of $10^{12}$ s$^{-1}$, we find a typical desorption rate of 6 ns$^{-1}$ for the smallest hydration shell, in fair agreement with our computational observations (see Fig. S8 and SI.S5). We highlight that ionic dynamics in the air following such desorption-mediated events cannot be characterized properly due to the finite size of our simulation box. The effective diffusion coefficient reported in Fig. 5D is accordingly evaluated by focusing on ionic nanodroplets in their adsorbed state. These observations are however a signature of the enhanced propensity for drops at low hydration numbers to detach and irreversibly evaporate from the surface (SI S5), a point we will come back to later.

We evidenced the occurrence of a peculiar transport mechanism for surface adsorbed ions on molecularly smooth hydrophobic surfaces, associated with an interfacially limited frictional transport process clearly distinct from that observed for ionic transport in bulk water. At a qualitative level, these conclusions are consistent with our two experimental observations of *(i)* mobility values which can exceed the standard bulk ionic diffusion coefficient for simple ions and *(ii)* a diffusion process independent of the exact ionic nature. The range of diffusion coefficients measured experimentally ($D \approx 1 - 7 \cdot 10^{-9}$ m$^2$.s$^{-1}$) falls within the lower range of our simulations, predicting $D \approx 3.10^{-9}$ m$^2$.s$^{-1}$ for a nanodroplet of 1.6 nm radius (Fig. 5D, upper axis). However, as reported above and discussed in Fig. S7, the experimental interfacial mobility exhibits slight sample-to-sample variations, suggesting the presence of additional surface-specific effects controlling interfacial ionic friction. These effects are possibly associated with heterogeneities in the substrate state at the molecular scale, due to defects in the grafted silane layer.

Within our simulation framework, ionic mobility is expected to show a strong dependence on hydration level. Experimentally, varying the ambient humidity would appear to be a simple way to tune the size of the nanodroplets and thus their effective friction with the solid surface. Going from low to high humidity levels, grand canonical ensemble simulations predict indeed an increase in the hydration number by a factor 20 for ions adsorbed at the surface, which would lead to a relative decrease of the diffusion coefficient by a factor 7 (Fig. S9). Unexpectedly, however, the ionic surface diffusion coefficient remains independent of the relative humidity in our experiments (SI Fig. S6). At this stage, several possible explanations can be invoked to rationalize this discrepancy.

First, the charge deposition process from moving contact lines is intrinsically out-of-equilibrium , and the initial hydration state of the nanodroplets on the surface is expected to be independent of ambient humidity (with a possibly large distribution in the size of the surface-deposited nanodroplets). Whether these surface-adsorbed ions might reach thermodynamic equilibrium over the course of our measurement remains unclear. In the case of bare silica surfaces, equilibration kinetics with vapor has been shown to reach up to several days[45]. In this situation, no clear effect of the ambient humidity would accordingly be expected to take place over the few hours of experiments carried out here.

In this respect, the transient charge relaxation kinetics evidenced at short times also appear unaffected by changes in ambient humidity (Fig. 3D and Fig. S6). Coming back to our numerical observations for the detachment of ions with low hydration levels, a speculative yet intringuing hypothesis is that this transient relaxation might stem from the early stage evaporation of the population of nanodroplets with very low hydration number. Such an effect would leave only the largest and bulkiest nanodroplets on the surface, diffusing at an effectively slower rate.

Finally, we cannot exclude the possibility of additional processes limiting and hindering surface transport compared to our numerical scenario of perfectly self-assembled hydrophobic monolayers. Such effects could stem for example from the interaction with localized surface defects, such as for example residual silanol groups. The presence of defects would effectively increase friction and interfacial retention compared to the idealized situation of a smooth silane layer assumed in our simulations, and could possibly weaken the dependence of transport on hydration or relative humidity, although the exact mechanism remains to be uncovered. Investigating how surface imperfections influence droplet retention and diffusion coefficients would accordingly be a valuable direction for future experimental and numerical studies.

**Conclusion**

Harvesting an in-situ scanning based potential mapping approach, we revealed the spatiotemporal evolution of surface-trapped charges deposited in the wake of droplets sliding down inclined hydrophobic surfaces. We showed that the deposited charge can be reversibly tuned by the bulk droplet pH, following similar trends as the surface charge born by non-ionizable hydrophobic surfaces. These observations point to the role of physisorbed ionic species (such as water self-ion and dissolved $CO_2$) in the surface charging mechanism. Remarkably, the charges initially deposited along the wake of the droplet tend to spread apart on the solid surface over time, consistent with a bidimensional diffusion process associated with ultra-high 2D gliding mobility exceeding that of standard salt in bulk water. Transport is furthermore found to be independent of the surface charge density, indicating that charge dynamics is dominated by surface interactions rather than inter-charge repulsions. We interpret our observations using Molecular Dynamics simulations, which reveal a peculiar mechanism for the transport of hydrated ions adsorbed on solid surfaces, uncorrelated to bulk transport properties and set by purely interfacial frictional interactions between the water molecules solvating these ionic charges and the solid surface.

Our observations and findings open up several exciting avenues and perspectives. In the context of solid electrification, the role of water adsorbates and mobile surface-adsorbed ions have been put forward to rationalize the charging of dielectric materials involving non-ionizable and ionizable polymer and hydrophobic materials[46–50], echoing the molecular picture presented here. Our measurements further stress that spatially and temporally resolved mapping are necessary to properly assess charge relaxation processes in liquid triboelectrification[20]. The dynamics of this new state of interfacial ionic matter further raises fundamental questions in the context of iontronics and nanofluidics. We can mention among others the role played by solvation on interfacial ionic friction[51], the possibility for electrostatic stabilization of molecular water layers at hydrophobic interface[52], the role of electronic interactions during droplet friction on conductive surfaces[6,53,54], up to the possibility for peculiar physical processes such as ionic pairing at these low-permittivity air interfaces[55]. The *ionic puddles* evidenced here have accordingly thus just started to reveal their secrets.

**Acknowledgments**

Z.B., E.V. and J.C. thank ESPCI PSL for funding the position of Z.B as a Teaching and Research Fellow. J.C. acknowledge funding from the ANR (grant 'GUACAmole' ANR-22-CE06-0003-01). This project has also received financial support from the CNRS through the MITI interdisciplinary programs and from the Carnot Institute 'IPGG Microfluidique'. S.G. acknowledges access to the HPC resources of IDRIS under the allocation 2023-AD012A14560 made by GENCI.

**Materials and methods**
Substrates consist of 2 mm thick borosilicate plates, which are made hydrophobic through liquid-phase silanization using a mixture of OctadecylTrichloroSilane (OTS) in toluene. To deposit charges on the surface, drops of an aqueous solution are launched from a 2 mm diameter nozzle and slide on the silanized surface, which is inclined at 67° with respect to the horizontal. Typically, 35 drops are launched for each experiment, allowing the system to reach a steady-state regime in droplet charge. Spatio-temporal surface potential maps are obtained by scanning an electrostatic voltmeter over the surface using two motorized stages. The voltmeter is a non-contact field compensation voltmeter, Trek 344-K-EX, associated with a probe 6000B-5C. The vertical distance between the probe and the glass surface is set to 0.7 mm, a value of the same order of magnitude as the probe diameter, which determines the approximate lateral resolution. The acquisition for one map lasts approximately 4 minutes.

**Supplementary Materials**
**Movie S1.** A droplet consisting of a single bicarbonate ion (red, yellow, and white beads) and 5 water molecules (transparent field) on a layer of $C_{18}$ molecules (dark and light beads). The movie represents a duration of 200 ps, with a time interval of 0.2 ps between consecutive frames.

**Movie S2.** A droplet consisting of a single bicarbonate ion (red, yellow, and white beads) and 161 water molecules (transparent field) on a layer of $C_{18}$ molecules (dark and light beads). The movie represents a duration of 200 ps, with a time interval of 0.2 ps between consecutive frames.

**SI Appendix.** Supplementary Information file.

# Supplementary Information

## *Giant mobility of surface-trapped ionic charges following liquid tribocharging*

Zouhir Benrahla[1], Tristan Saide[1], Louis Burnaz[1], Emilie Verneuil[1],
Simon Gravelle[2], Jean Comtet[1]

[1] *Soft Matter Sciences and Engineering, CNRS, École supérieure de Physique et de Chimie Industrielles de la Ville de Paris, Université Paris Sciences et Lettres, Sorbonne Université, Paris 75005, France*
[2] *Univ. Grenoble Alpes, CNRS, Laboratoire Interdisciplinaire de Physique,38000 Grenoble, France*

### S1. Surface preparation

Here, we describe the procedure to obtain large hydrophobized glass plates through a silanization process with Octadecyltrichlorosilane (OTS).
The borosilicate glass plates of size 180x80 mm and thickness 2 mm (*Ediver*, Borofloat 33), were cleaned with distilled water, acetone, and dried under a flux of nitrogen. The glass plates were then treated with piranha (a solution of 70% sulfuric acid $H_2SO_4$ and 30% hydrogen peroxide $H_2$) to eliminate organic traces on the surface. The plates were immersed in this solution at 200°C for 20 min, then left for 1 hour, without heating. Once the temperature had dropped sufficiently, the plates were soaked in a deionized water bath, followed by 2 ultrasonic baths in deionized water, and finally dried with nitrogen. The surface was then activated with UV/ozone for 15 minutes and immersed for 15 min in a volume of anhydrous toluene with 0.25% of OTS in volume. The plate was then sonicated in a toluene bath for 15 min and dried with nitrogen.

To characterize the surface state of the silanized glass plates, we measured the contact angle hysteresis, defined as the difference between the advancing ($\theta_a$) and receding ($\theta_r$) angle of a water drop deposited on the substrate. We find $\theta_a \approx 115°$ and $\theta_r \approx 45°$, with a contact angle hysteresis of 70 °.
The homogeneity and thickness of the OTS layer were characterized using ellipsometry on reference silicon wafers submitted to the same chemical process as the glass plates. The silanized thickness was homogeneous at the wafer scale, within an experimental lateral resolution of ca. 1 $\mu m^2$. We found a grafted thickness of 2.3 nm, consistent with literature values [1].

These ellipsometry measurements suggest the occurrence of a homogeneous surface state at the micrometer scale, although we cannot rule out the presence of localized chemical heterogeneities, which could be responsible for the contact angle hysteresis observed in our samples.

In order to neutralize the electrostatic charges at the surface between experiments, the plates were sonicated in a distilled water bath for 15 min and dried with nitrogen. The surfaces were placed in an oven at 150 °C for 45 min, in a glass petri dish covered with aluminum foil. This procedure led to a surface state characterized by a homogeneous surface state free of residual surface charges, showing RMS voltage lower than 2 V (Main text, Fig. 2A).

## S2. Experimental setup

**Surface charging through droplet sliding**. During a typical experiment, drops of aqueous solution are launched from a 2 mm diameter nozzle (Fig. S1, in blue) and slide on the silanized surface, inclined at 67° with respect to the horizontal with an approximate velocity of 55 $cm.s^{-1}$. The water drops drip from a nozzle fed by a syringe pump at a rate of 2 drops per second, and are grounded upon detaching from the nozzle. Drops have an approximate radius of 1.9 mm and fall onto the surface from about 13 mm. The bottom of the glass plate is in contact with an aluminum layer and grounded. Typically, 35 drops are launched for each experiment, allowing to reach a steady-state regime in droplet charge, as shown in Figure S2. The sliding droplet motion is recorded using a Photron Fastcam UX100 ultra-fast camera with an acquisition rate of 1000 fps. These videos allow us to check that the drops do not deviate upon successive slides and match their wake to the charge distribution (see Fig. S3).

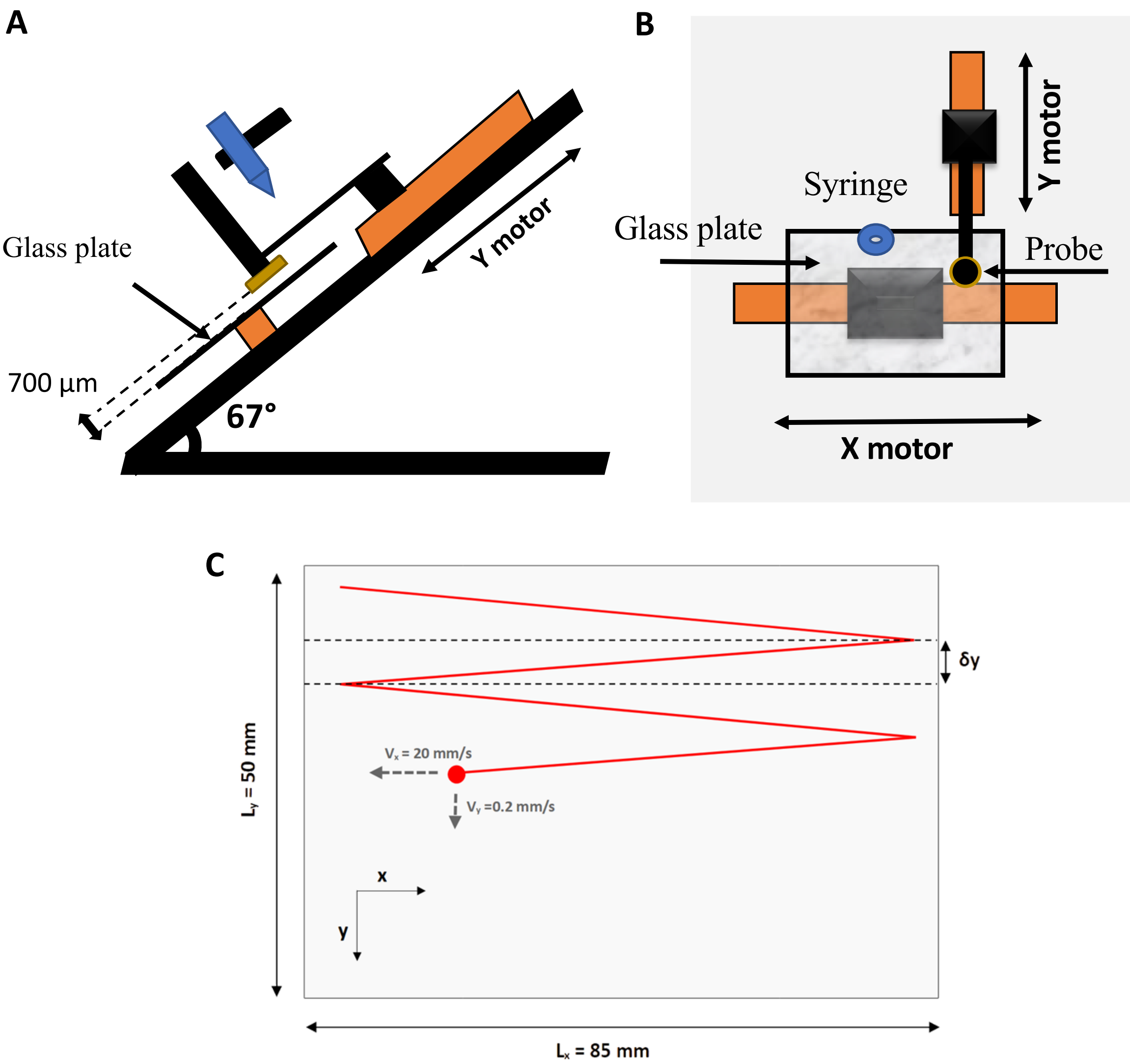

**Figure S1. Experimental set-up.** 2D schematic of the experimental setup in side **(A)** and top **(B)** view. The set-up is further placed in a Faraday box (not represented in the figure) to minimize electromagnetic noise and to allow for environmental control. **(C)** Diagram of the potentiometer probe path over the plate. Both plates operate simultaneously to reduce scan time, so the path takes a saw-like shape. The vertical offset between each line is $\delta y = v_y \cdot L_x = 0.85$ mm. This value is of the same order of magnitude as the spatial resolution of the probe, which justifies to neglect $\delta_y$ in front of $L_x$ and $L_y$ when processing the data.

**Operating principle of the electrostatic voltmeter.** Surface potential maps were obtained using a non-contact field compensation voltmeter trek 344-K-EX associated with a probe 6000B-5C, which features a disk-shaped probe of 0.79 mm diameter.

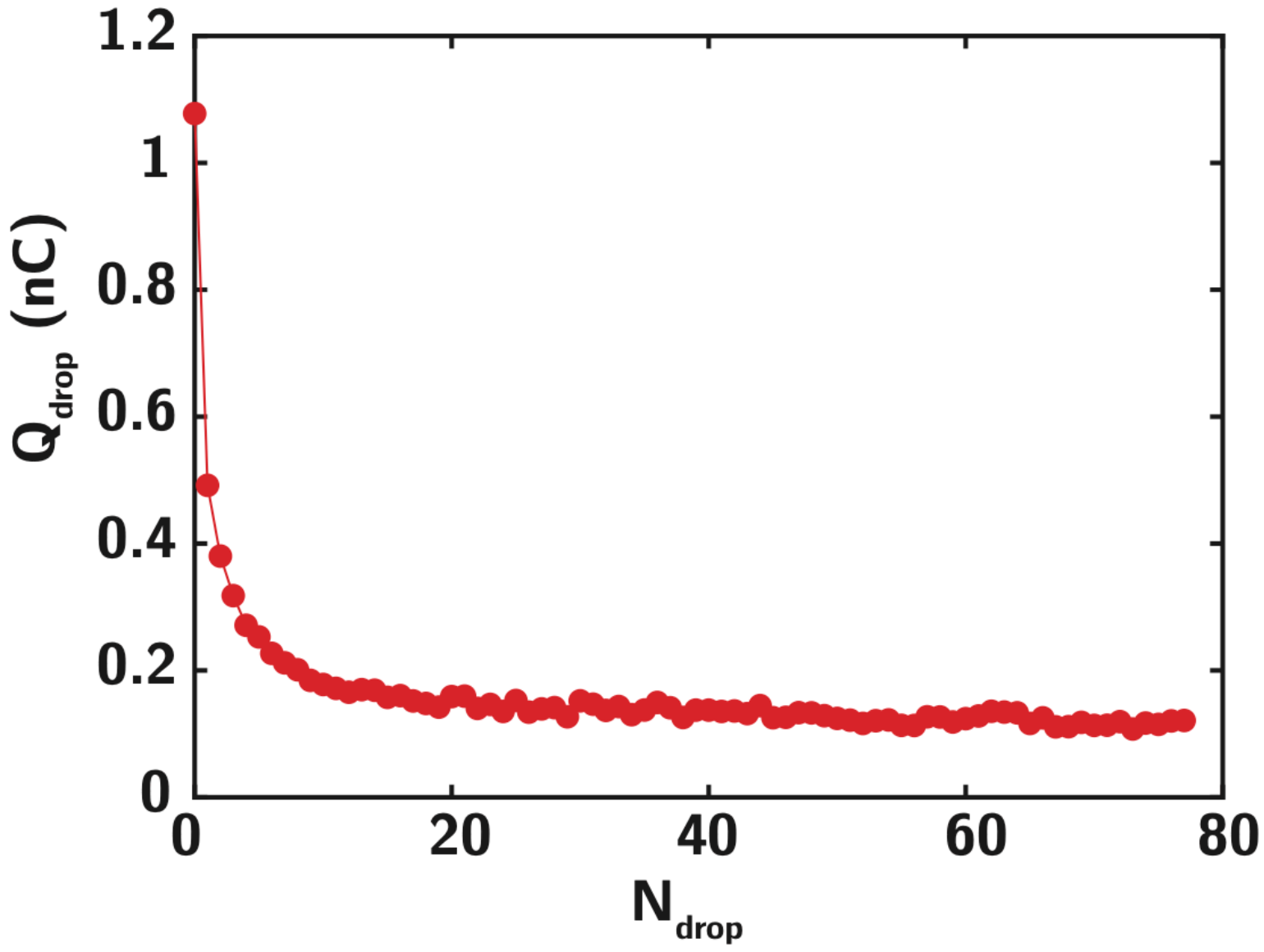

**Figure S2. Drop charging.** Evolution of the droplet charge $Q_{\mathrm{drop}}$ as a function of the number of sliding droplets. Drop charge is measured by integrating the discharge current following droplet surface sliding between two electrodes, using a similar setup as described in [2]. In this experiment, the surface is a silanized Si/$SiO_2$ wafer (WAP2FZ from Neyco). The native $SiO_2$ oxide layer of the wafer leads to a similar surface chemistry as the glass plates and similar OTS grafting. The wafer is undoped, with a resistivity larger than 20 000 $\Omega.\mathrm{cm}^{-1}$. The surface is inclined at 45° with respect to the horizontal (the inclination angle makes weak effects on drop charging). The interelectrode distance is $L = 22$ mm and the time interval between sliding droplets is 3.3 s.

**Potential mapping.** A schematic of the experimental setup for mapping is presented in Fig. S1. Briefly, the electrostatic voltmeter (yellow disk) is attached to an XYZ micro positioning stage and placed above the glass surface. The vertical distance between probe and glass surface is set to 0.7 mm, a value of the same order of magnitude as the probe diameter, setting the approximate lateral resolution. The position of the probe along $y$ is controlled using a MTS50-Z8 motorized stage (Thorlabs), while the position of the plate along $x$ is controlled using a DDS100 motorized stage (Thorlabs).
During scanning, the probe moves periodically back and forth over 50 mm along the Y direction with a speed of 0.2 $mm.s^{-1}$ while the glass slide moves periodically back and forth over a distance of 85 mm along the X direction with a speed of 20 $mm.s^{-1}$ (Fig. S1C). The voltmeter signal and motor displacements are acquired using a NI card DAQ (NI USB-6216), with an acquisition frequency of 2 kHz, and the signal further processed using a home-made matlab code. The acquisition for one map lasts for nearly 4 minutes (~251 seconds).

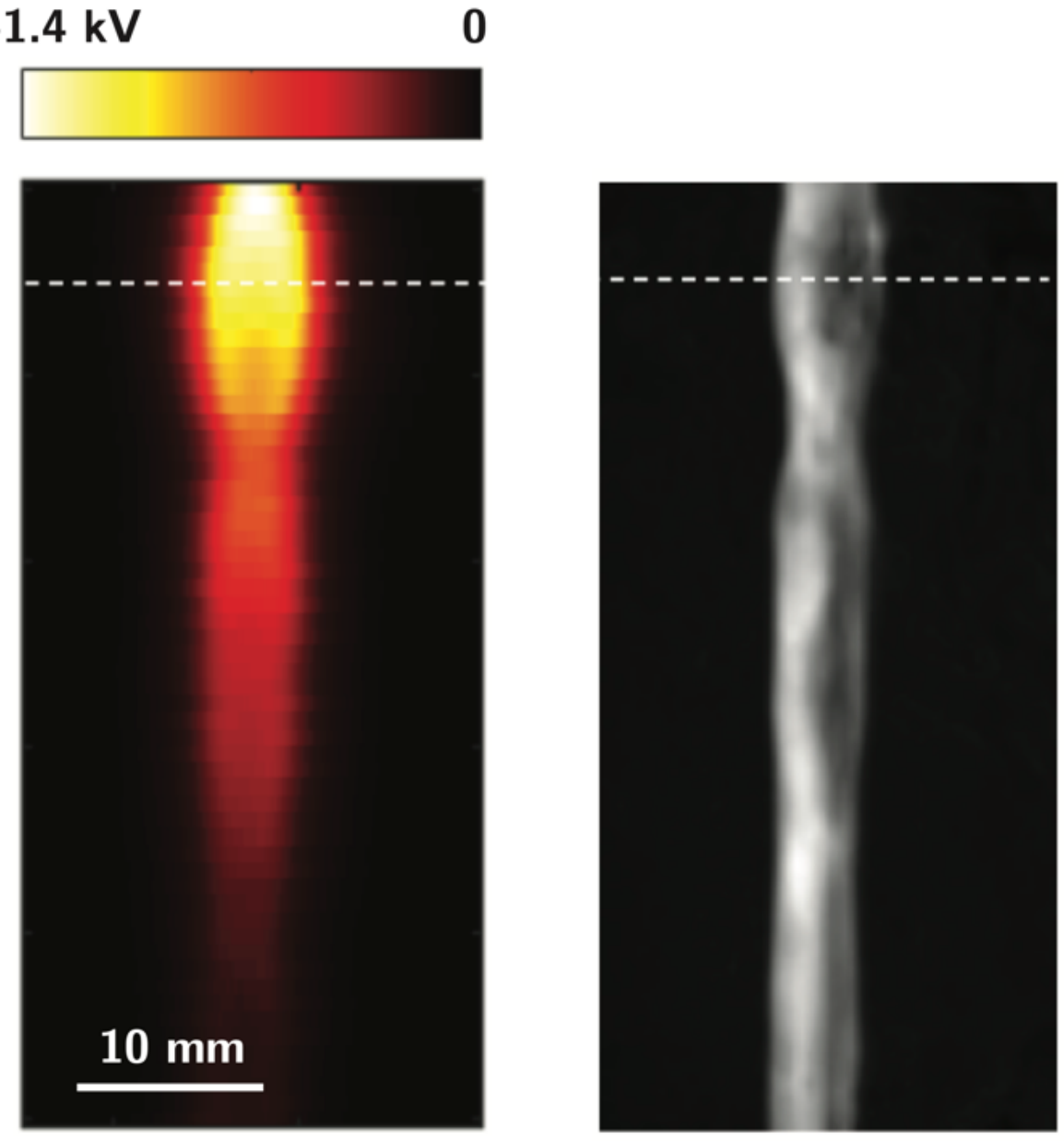

**Figure S3. Charging and drop path.** Comparison of the charge distribution (left) after 2.2 minutes with the 35-drop path envelope (right), obtained by displaying the high-speed video's standard deviation. The width at half-height $\Delta$ of the potential distribution for y = 5.5 mm (dotted line) is similar in magnitude to the maximum width of the envelope obtained from the videos ($\Delta = 6.7$ mm versus 6.3 mm), consistent with the fact that surface charges are deposited along the wake of the drop.

## S3. Data processing for surface diffusion

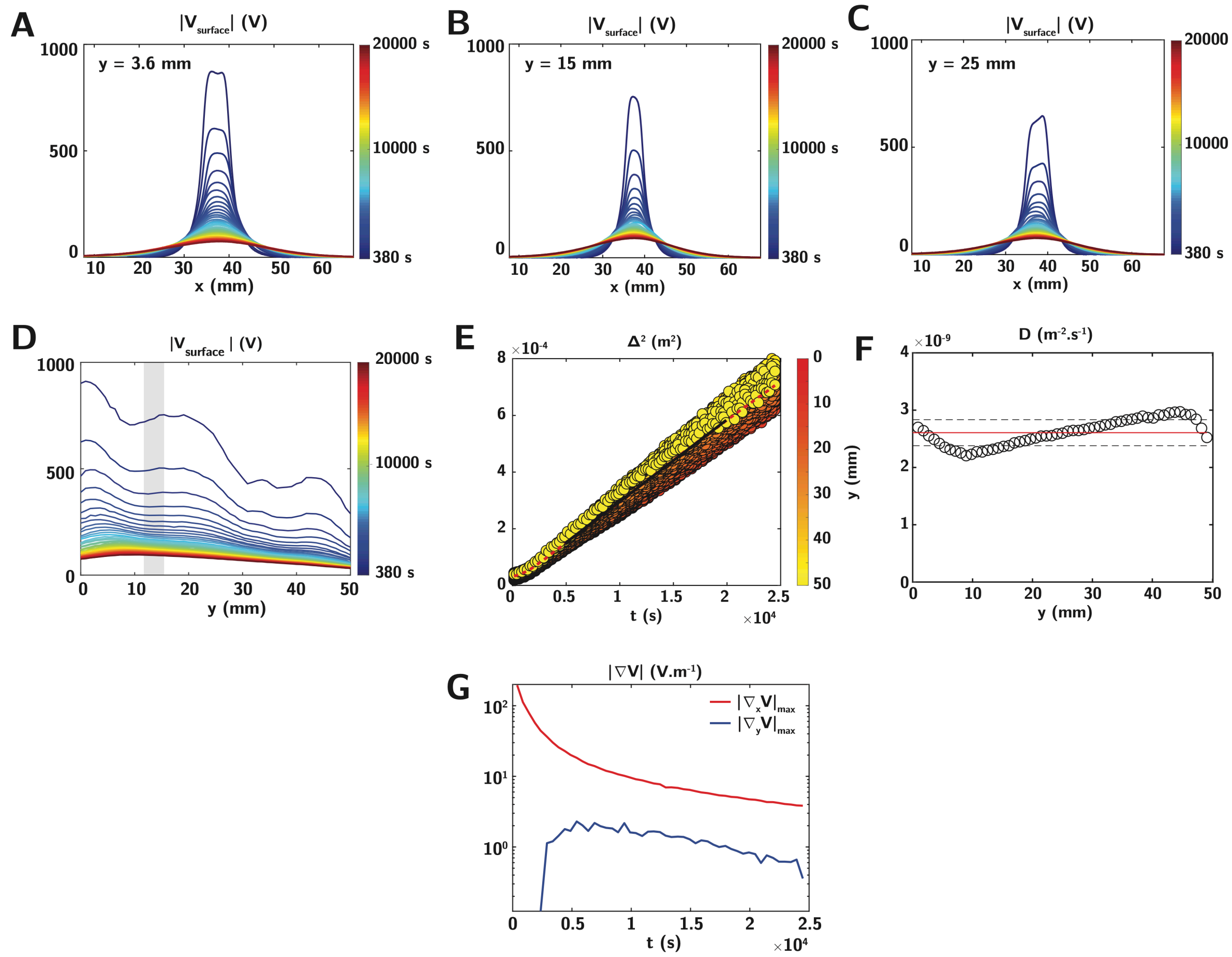

**Figure S4. Quantification of the diffusion process. (A-C)** Transverse potential profile along $x$ for various longitudinal positions $y \in [3.6, 15, 25]$ mm along the wake of the drop and time ranging from 380 to $2.10^4$ seconds. **(D)** Evolution of the potential profile along the vertical $y$ direction, in the center of the drop wake, for time ranging from 380 to $2.10^4$ seconds. **(E)** Evolution of the squared FWHM $\Delta^2$ with time for various positions in $y$ from 0 to 50 mm. The red dashed line represents the average of $\Delta^2$. The black line is a linear fit showing a diffusion coefficient $D = 2.6 \cdot 10^{-9}$ m$^2$.s$^{-1}$. **(F)** Evolution of the diffusion coefficient $D$ calculated at various positions $y$ from (F). The red dashed line is the average of $D$ and the dotted lines represent $\pm 1$ standard deviation. **(G)** Temporal evolution of the steepest potential gradient along the longitudinal ($|\nabla_x V|_{\max}$, red) and vertical ($|\nabla_y V|_{\max}$ blue) directions in the region $y \in [12 - 15]$ mm (D, grey dashed box). $|\nabla_x V|_{\max}$ is taken at the inflexion point of the gaussian-like profile along $x$. $|\nabla_y V|_{\max}$ is evaluated as the local slope of the potential along the vertical direction $y$.

We detail here our approach for data processing, considering the data shown in Fig. 3 of the main text. Looking at surface potential profiles along the horizontal transverse direction along $x$, we observe at short times a gate-like charge distribution (Fig. 3B and Figs. S4A-C), with a lateral extension fixed by the drop sliding path (Fig. S3). Over increasing time, as diffusion intervenes, this initial shape tends towards a Gaussian-like profile. As shown in Fig. S4D, along the longitudinal y-direction (i.e. along the sliding path of the drop), the surface potential at short time shows an overall decrease along the sliding path (blue curve), with slight oscillations. These local variations are smoothed out by diffusion, and the profile quickly reaches a state of small local potential gradients $\nabla_y V$.

Given these observations for the shape of the charge distribution, we approximate diffusion by a 1D process along the transverse $x$ direction. As these transverse profiles along $x$ cannot be properly approximated as gaussian at short times, we characterize instead the temporal evolution by their squared Full Width at Half Maximum $\Delta$, related to the width $\sigma$ of the Gaussian profile observed at long time by $\Delta = 2\sqrt{2\ln(2)} \cdot \sigma$. Following the linear temporal evolution $\sigma^2 = 2Dt$, the local diffusion coefficient $D$ can be simply extracted from the temporal evolution of $\Delta$, as $\Delta^2 = 16\ln(2) \cdot Dt$. As shown in Fig. S4E, $\Delta^2$ is evaluated along various positions in $y$, and shows a clear linear increase with time, consistent with a purely diffusive process. The various FWHM show limited variations along $y$ (going from red to yellow in Fig. S4E). We can further evaluate the local diffusion coefficient $D$ along these various positions, as shown in Fig. S4F. As again evidenced here, $D$ is approximately independent of $y$, justifying our treatment of the diffusive process as purely 1D and the transverse $x$ direction. Finally, this approximation of a 1D diffusive process is further validated by evaluating the potential gradients along the transverse ($x$) and longitudinal ($y$) directions (Fig. S4G), showing that $|\nabla_x V| \gg |\nabla_y V|$ for all time.

## S4. Physicochemical effects for charge diffusion and relaxation

Most experiments were carried out using deionized water. To probe the effect of pH, aqueous solutions were prepared through the addition of KOH and HCl to a solution with a background leve of 10 mM NaCl.

**Charge relaxation.** The charge deposited on the surface is calculated from the integral of the local surface potential $V_S$, as $Q_S = \epsilon\epsilon_r/h \cdot \iint_A V_s(x,y)dx.dy$, where $h = 2$ mm in the glass thickness, $\epsilon_r \approx 4.8$ the relative dielectric constant of the glass, $\epsilon_0$ [F.m$^{-1}$] the vacuum permittivity, and the integral runs over the scanning area $A \approx 85x50$ mm$^2$ (Cf. Fig. S1C).
In order to verify that the charge decay observed at long timescales is not due to charge leakage out of the scanning area of finite size, we compute the integrated charge in a sub-region where the vertical potential gradients along $y$ on the top and bottom boundaries are approximately equal (Fig. S5C, inset).

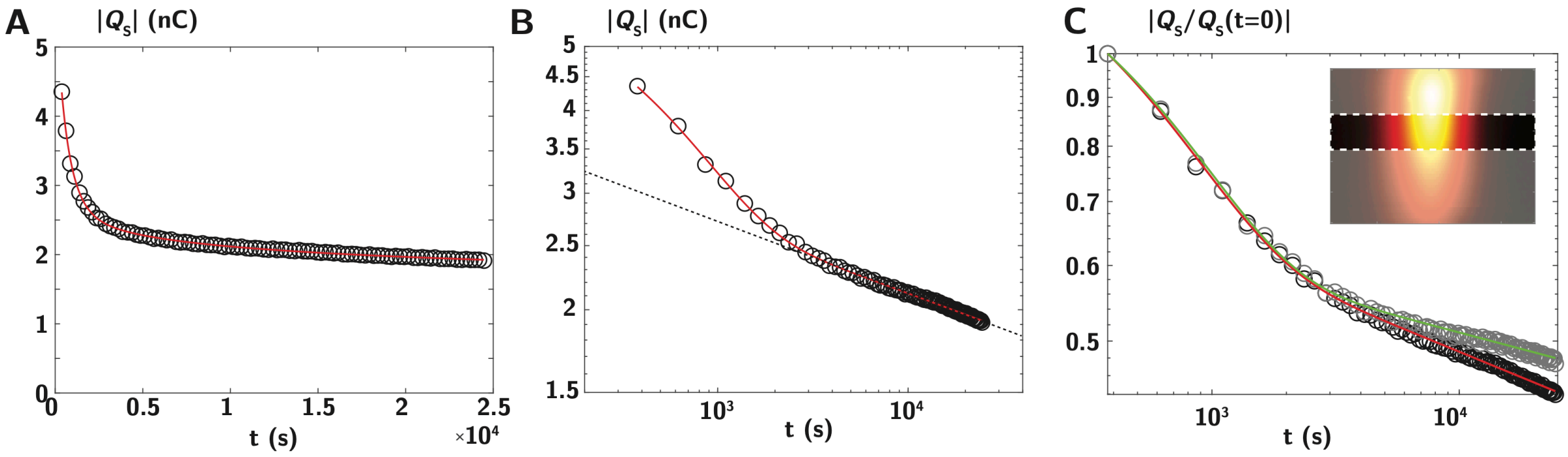

**Figure S5. Temporal relaxation of the charge deposited on the surface $Q_S$. (A)** Evolution of the integrated deposited surface charge $Q_S$ with time. The red line is an empirical fit $Q_S(t) = Q_1 \cdot \exp\left(-\frac{t}{\tau}\right) + Q_2 \cdot t^{-\alpha}$, characterizing the temporal evolution of the total charge in the relaxation regime by the combination of a transient exponential process with a slow power-law decay. **(B)** Same figure in logarithmic coordinates. The dashed black line shows a power-law scaling. We find in our various experiments typically $\tau \approx 500 - 1500$ s $\approx 10 - 25$ minutes and $\alpha \approx 0.1 - 0.6$. **(C)** Comparison of the relaxation of the normalized surface charge $|Q_S/Q_S(t=0)|$ evaluated over the entire image (red) and in a subregion (white box in inset), in which the vertical gradient along $y$ on top and bottom are approximately equal (green). The charge in this region shows a similar power-law decay with time, suggesting that the decay is indeed due to charge relaxation and not to charge leakage outside of the observation area.

## Role of humidity

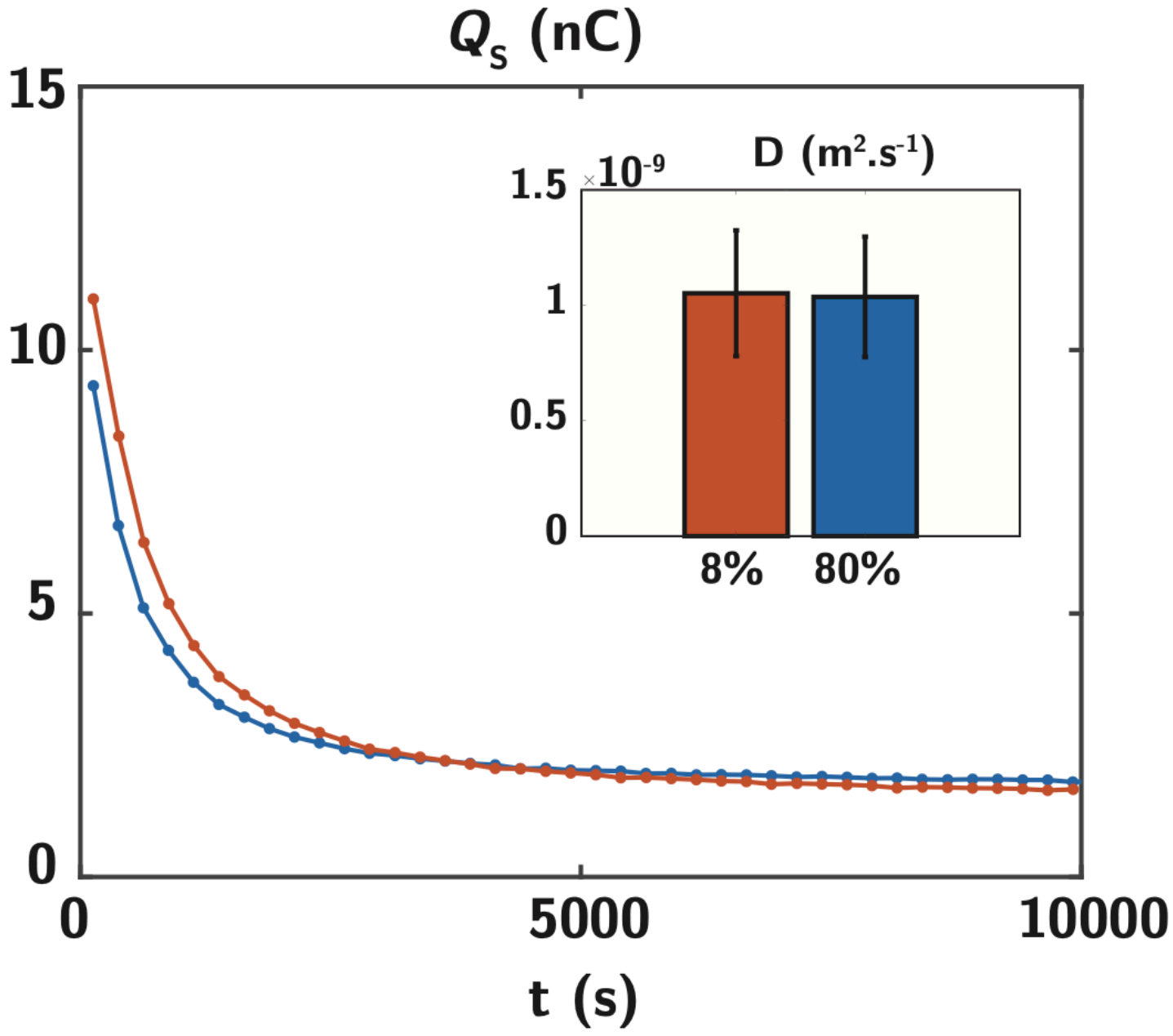

**Figure S6. Role of humidity.** We probed the role of humidity on the charge diffusion process. To carry out experiments at high relative humidity, we placed a small beaker filled with water inside the chamber. For low-humidity experiments, nitrogen was continuously flushed into the chamber. The evolution of chamber humidity and temperature was measured systematically for each experiment by a ROTRONIC probe. The figure shows the evolution of the charge relaxation of the surface $Q_s(t)$ with time at high (blue, $RH = 80\%$) and low (red, $RH = 8\%$) humidity. The inset shows the associated diffusion coefficient, which appears independent of $RH$.

## Surface to surface variations

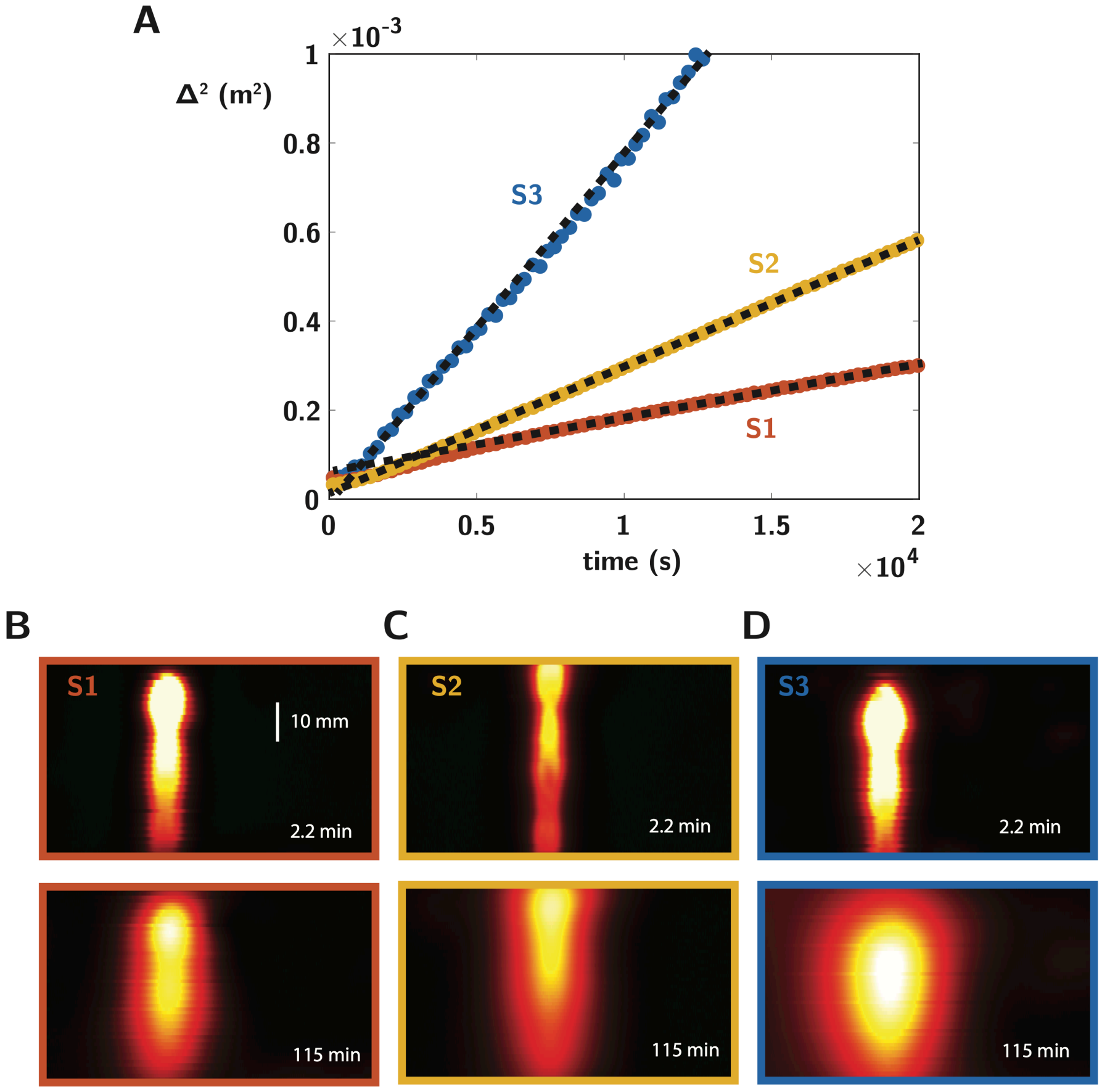

**Figure S7. Surface to surface variations. (A)** Evolution of the Full Width at Half Maximum squared $\Delta^2$ with time on three representative surfaces S1, S2 and S3, showing an effective diffusion coefficient $D_1 \approx 1.10^{-9}$ $m^2.s^{-1}$, $D_2 \approx 2.6 \cdot 10^{-9}$ $m^2.s^{-1}$, $D_3 \approx 7 \cdot 10^{-9}$ $m^2.s^{-1}$ (respectively in orange, yellow and blue). Black dashed lines are linear fits. **(B-D)** Evolution of the surface charge profile at initial time (2.2 min) and later time (115 min) on the three surfaces. For clarity, the color scale in adjusted to maximize contrast. The conditions for S1 correspond to a solution with 10 mM NaCl and the conditions for S2 and S3 correspond to DI water.

## S5. Details for the Molecular Dynamics simulations

Molecular dynamics simulations of a single hydrated ion on a self-assembled monolayer (SAM) surface were performed using LAMMPS [3]. In total, three types of ions were simulated, $OH^-$, $H_3O^+$, and $HCO_3^-$. Ions were solvated with a number $N_w$ of water molecules ranging from 5 to 249. A SAM surface with an area of 3x3 nm$^2$ was made of 154 $C_{18}$ molecules grafted on the free silicon atoms of a $Q^3$ silica surface, corresponding to a surface density of $C_{18}$ molecules of 5 nm$^{-2}$. The final average thickness of the SAM layer was measured to be 2 nm using Gibbs dividing plane analysis to detect the interfaces of the layer. To reduce computational cost, the silica surface was not explicitly modeled, and the $C_{18}$ molecules were anchored at the positions of the free silicon atoms instead. The simulation box was 7.5 nm in the *z* direction normal to the SAM surface, and reflecting walls were used in this direction. Periodic boundary conditions were used along the *x* and *y* directions. Long range Coulomb interactions were treated using the particle-particle/particle-mesh method extended for non-neutral slab systems [4]. A cut-off of 1.2 nm was used for the Lennard-Jones potential.

The TIP4P/2005 model was used for water [5], the GROMOS force field was used for the $C_{18}$ molecules and $HCO_3^-$ ions [6], and the parameters from Bonthuis et al. were used for the $OH^-$ and $H_3O^+$ ions [7]. Ions and water molecules were treated as rigid, while the $C_{18}$ molecules were treated as flexible. A timestep of 1 fs was used.

**MSD and drop jumping rate measurements.** For every simulation, a single ion was initially placed near the SAM surface, and a number $N_w$ of water molecules were randomly placed in a half-sphere centered on the ion. The system was relaxed for 100 ps, and then the position of the ion was recorded for 500 ps. For each configuration, 30 statistically independent simulations were performed. The temperatures of the water molecules and $C_{18}$ molecules were maintained at 300 K using two Nosé-Hoover thermostats [8], [9]. The mean squared displacement (MSD) of the ion was then recorded along the *x* and *y* directions.

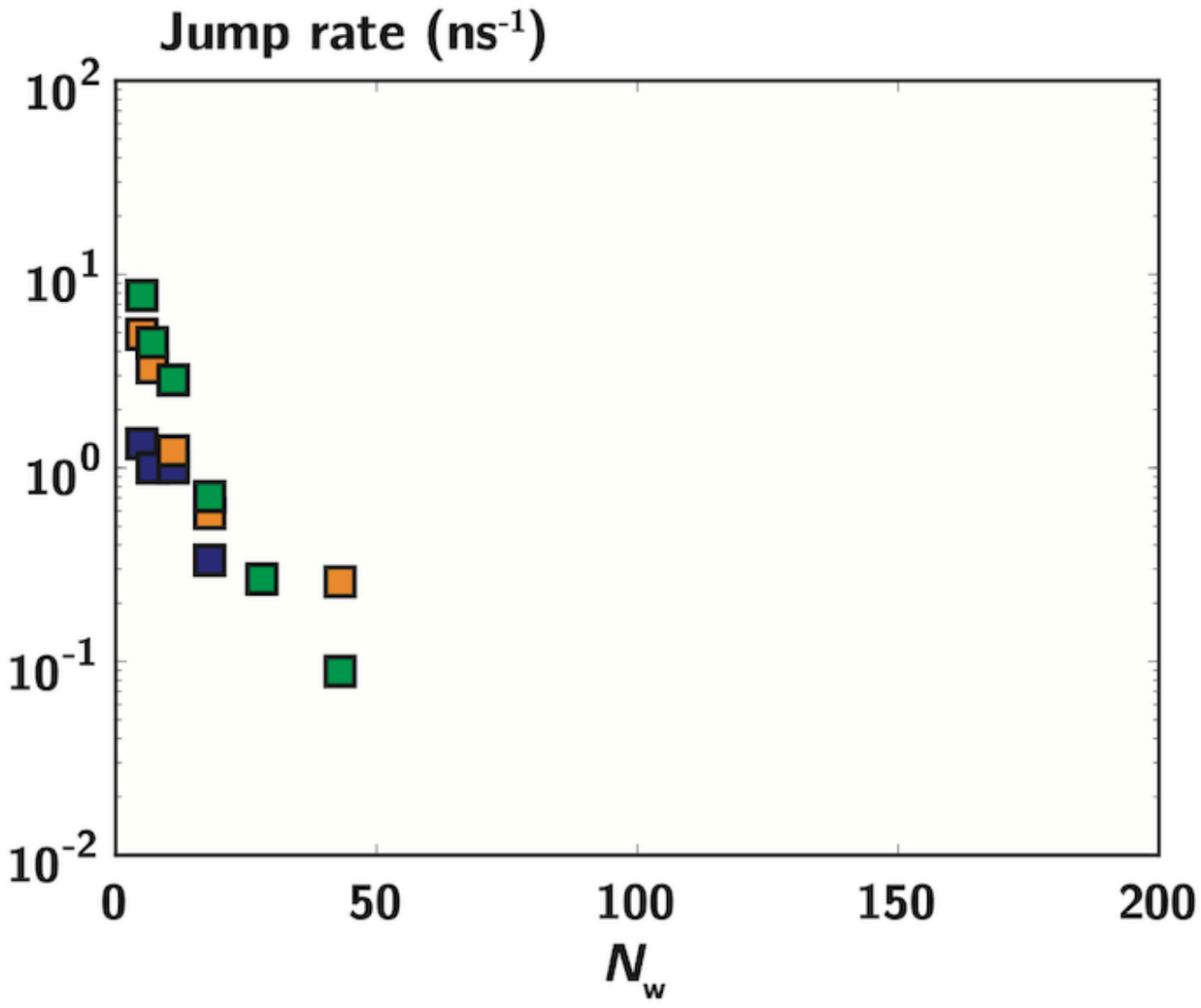

**Figure S8. Jump rate for the ionic droplets as a function of the number of adsorbed water molecules** (blue, bicarbonate; orange, hydronium; green, hydroxide). The jump rate could not be evaluated over the restricted simulation time for $N_w > 50$. Considering an adsorption energy $U_{ads} = 5k_BT$ for $N_w = 5$ (Fig. 5D) and taking an attempt frequency $f_0 = 10^{12}$ s$^{-1}$, we get a desorption rate $k = f_0 e^{-U_{ads}/k_BT} = 6.710^9$ s$^{-1}$, close to the numerically measured rate of 1.33 10$^9$ s$^{-1}$ in these conditions.

**Droplet jumps and evaporation rate.** The droplet jumps observed for low hydration number correspond to transient evaporation from the surface, with a return rate that could be described from first passage time theory and a return probability to the surface expected to reach unity over long observation timescales. In actual experimental conditions, such transient desorption might lead to a net evaporation out of the surface, yet with a rate much smaller that the effective desorption rate.

**Grand Canonical Monte Carlo / Molecular Dynamics (GCMC/MD) Simulations.** To predict the number of water molecules per ion at a given water chemical potential, hybrid Grand Canonical Monte Carlo / Molecular Dynamics (GCMC/MD) simulations were performed using the LAMMPS simulation package. The initial configuration consisted of either a single ion in vacuum or a single ion near a SAM surface.

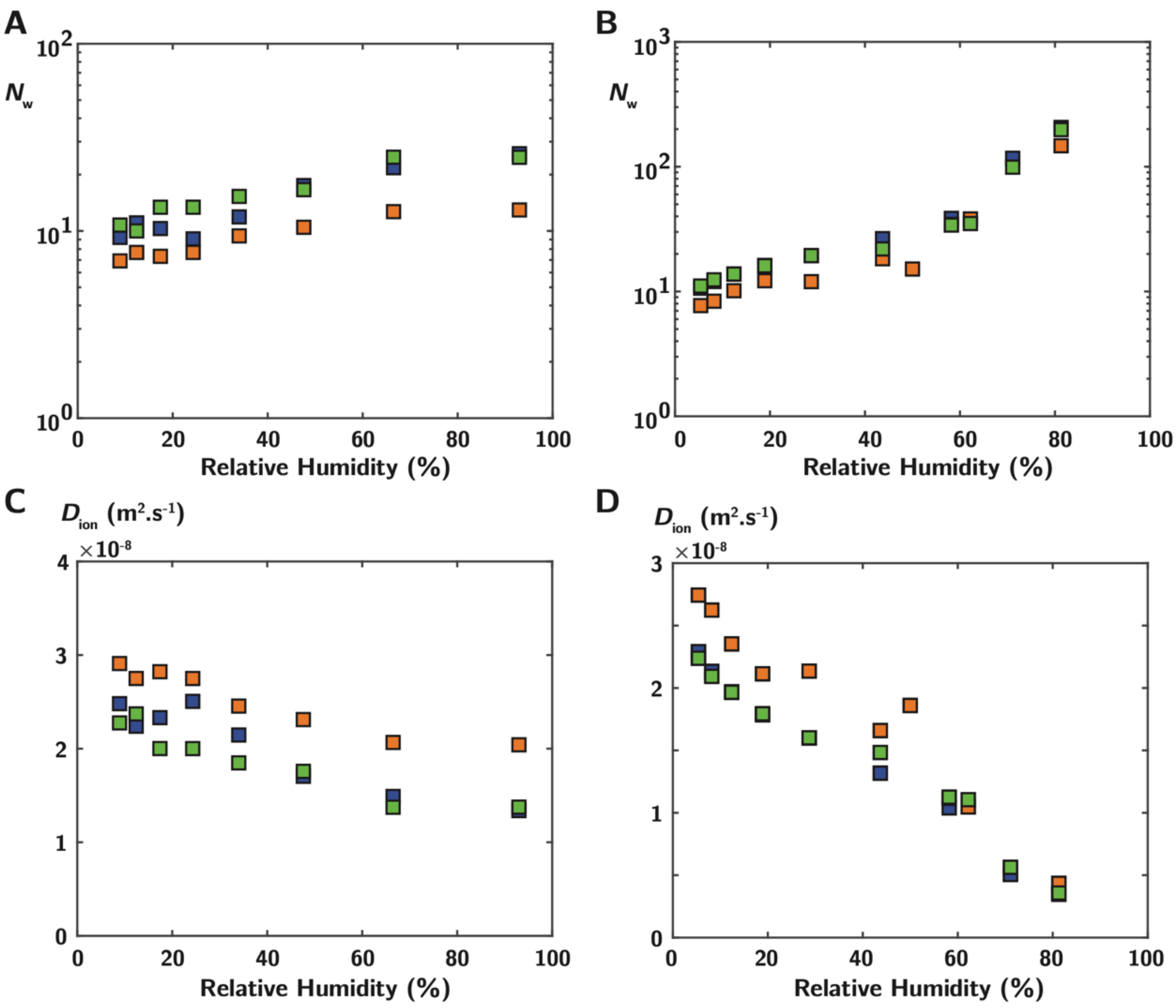

**Figure S9: Effect of ionic nature on adsorbed water and ionic surface diffusion**. Number of adsorbed water molecules as a function of relative humidity for **(A)** isolated single ions and **(B)** ions at a SAM surface, comparing bicarbonate (blue); hydronium (orange) and hydroxide (green) as obtained from GCMC/MD simulations. **(C-D)** Evolution of surface diffusion coefficient as a function of relative humidity for surface-adsorbed ions surrounded by a number of water molecules estimated for **(C)** isolated single ions and **(D)** ions at a SAM surface (see A and B). The relation between the diffusion coefficient and the number of adsorbed water molecules is taken from the geometrically limited dashed fit in Fig. 5D, with an interfacial friction coefficient $\lambda = 1.8 \cdot 10^5$ Pa.s.m$^{-1}$.

During the GCMC steps, water molecules were inserted or deleted based on a specified chemical potential μ corresponding to an external fictitious reservoir ($\mu \sim RT\ln(RH)$, where $RH$ is the relative humidity, $R$ the gas constant, and $T$ the temperature). In the MD steps, atomic positions were relaxed under the $NVT$ ensemble (constant number of atoms $N$, constant volume $V$, and constant temperature $T$). These GCMC and MD steps were carried out alternately until the number of water molecules converged to a plateau (see results in Fig. S9).

**Free energy profile measurements.** Free energy profile measurements were performed to probe the interfacial physisorption energy (Fig. 5E). The free energy profile of the droplet was reconstructed as a function of the coordinate $z$, normal to the SAM surface using the weighted histogram analysis method (WHAM). Umbrella sampling was employed with windows spaced by 3 Å. For $N_{\mathrm{w}} = 249$, a harmonic biasing potential with a force constant of $k = 2.5$ kcal.mol$^{-1}$ was used, while for $N_{\mathrm{w}} = 5$, a lower force constant of $k = 0.5$ kcal.mol$^{-1}$ was applied.

**Slip length measurements.** Slip length measurements were performed using a Couette flow setup. The system consisted of two 3 × 3 nm² SAM surfaces confining a layer of pure water, represented by 1300 water molecules. Translational motion was applied to each SAM surface at a velocity of 20 m/s. The position of the water-SAM interface was determined based on Gibbs dividing plane analysis, and the slip length was then estimated by fitting the velocity profile. The slip length $b_{\mathrm{sim}}$ was found to be approximately 1.6 nm.

**Contact angle measurements and droplet contact area.** Contact angle measurements were performed using the same protocol described in Ref. [10]. A pure water droplet consisting of 2500 water molecules was placed on a SAM surface with lateral dimensions of 6 × 6 nm$^2$. The system was equilibrated at a temperature of 300 K for 50 ps, and the density was then recorded for 2 ns. The contact angle was estimated by fitting a straight line to the density profile at $y \rightarrow 0$, and was found to be around 90°, in good agreement with the experimental measurements on macroscopic water droplets.
Assuming liquid water is incompressible, a time-averaged droplet radius $R$ can be estimated from the number of water molecules $N_{\mathrm{w}}$ as $R \approx \left[\frac{3.N_{\mathrm{w}}.V_{\mathrm{w}}}{2\pi}\right]^{1/3}$, where $V_{\mathrm{w}}$ is the molecular volume of water. The corresponding surface area in contact with the substrate then scales as $A = \pi \times R^2 = \pi \times \left[\frac{3.N_{\mathrm{w}}.V_{\mathrm{w}}}{2\pi}\right]^{\frac{2}{3}}$.

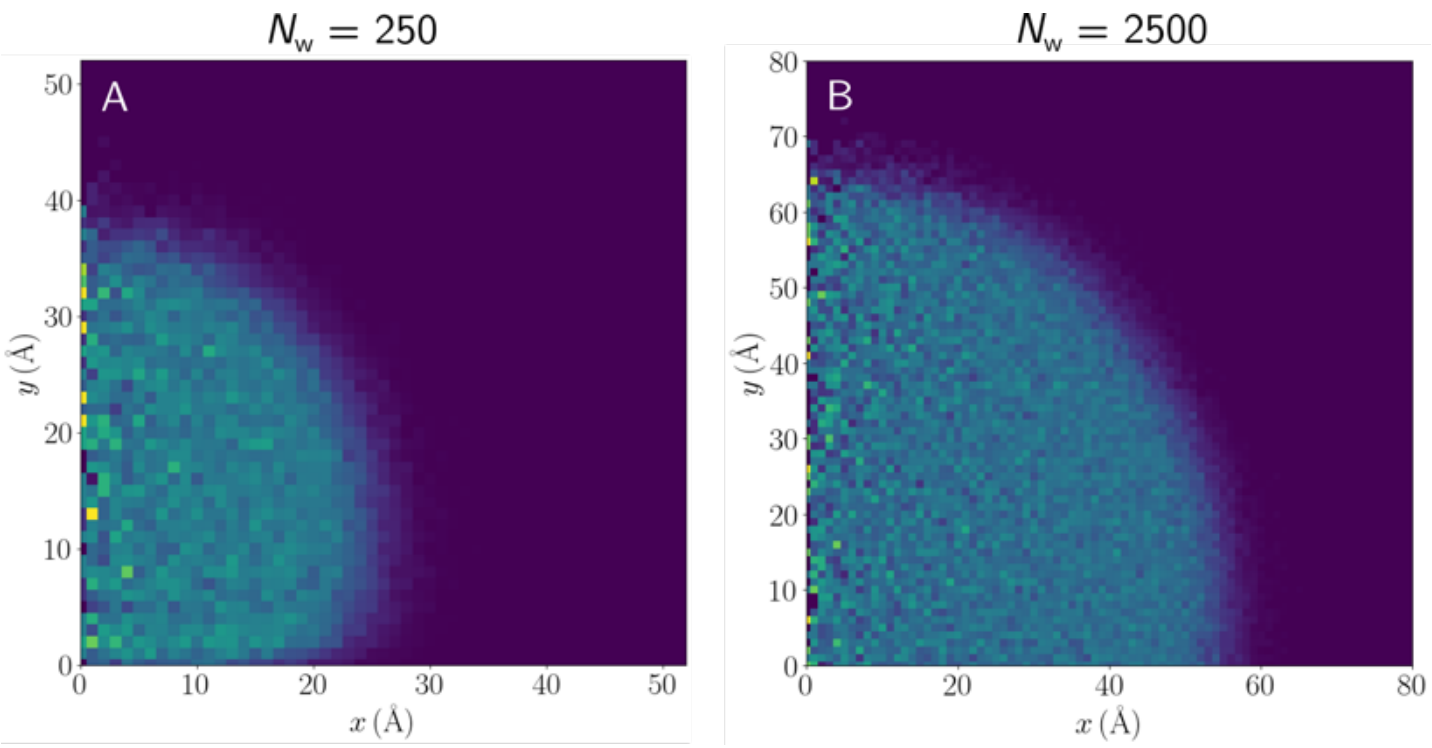

**Figure S10. Time-averaged density profiles** for droplets with $N_w = 250$ (left) and 2500 (right). Light blue regions represent high-density areas, while dark blue indicates low-density regions. The interface with the SAM is positioned at $y = 0$, as determined through Gibbs dividing plane analysis. The position $x = 0$ corresponds to the droplet's center of mass.